 \documentclass{aa}  

\usepackage{graphicx}
\usepackage{txfonts}
\newcommand{\xmm}{{\em XMM-Newton}}
\newcommand{\chandra}{{\em Chandra}}
\newcommand{\rhoOph}{$\rho$~Ophiuchi}
\newcommand{\fxu}{{erg~s$^{-1}$~cm$^{-2}$}}

\begin{document} 
\title{The first stars of the {\em Rho Ophiuchi Dark Cloud.}}
  \subtitle{XMM-Newton view of Rho Oph and its neighbors.} 

\author{I. Pillitteri\inst{1,2}
\and
S. J. Wolk\inst{2}
\and
H. H. Chen\inst{2}
\and
A. Goodman\inst{2}
}

\institute{INAF-Osservatorio Astronomico di Palermo, Piazza del Parlamento 1,
    Palermo, 90147 -- Italy
              \email{pilli@astropa.inaf.it}
         \and SAO-Harvard Center for Astrophysics, 60 Garden St, 
              Cambridge, 02138, MA -- US
}

   \date{Received; accepted}

\abstract{ Star formation in molecular clouds can be triggered by the dynamical action of winds from massive stars. 
Furthermore, X-ray and UV fluxes from massive stars can influence the life time of surrounding circumstellar disks.
We present the results of a 53 ks \xmm\ observation centered on the \rhoOph~A+B binary system.  
\rhoOph\ lies in the center of a ring of dust, likely formed by the action of its winds. 
This region is different from the dense core of the cloud (L1688 Core F) where star formation is at work.
X-rays are detected from \rhoOph\ as well as a group of surrounding X-ray sources. 
We detected 89 X-ray sources, 47 of them have at least one counterpart in 2MASS+{\em All-WISE} catalogs.
Based on IR and X-ray properties, we can distinguish between young stellar objects (YSOs)
 belonging to the cloud and background objects. 
Among the cloud members, we detect 3 debris disk objects and  
22 disk-less $-$ Class III  young stars. 
{ We show that these stars have ages in $5-10$ Myr, and  are significantly 
older than the YSOs in L1688. We speculate that they are the result of an early burst 
of star formation in the cloud.}  An X-ray energy of $\ge5\times10^{44}$ ergs has been 
injected into the surrounding medium during the past $5$ Myr, we discuss the effects of 
such energy budget in relation to the cloud properties and dynamics.
}

   \maketitle
%

\section{Introduction}
The action of an external agent onto an initial gas cloud is invoqued
to create gas instabilities and lead to the formation of the protostellar cores  
\citep{Elmegreen1998}. While supernova explosions are one possible agent, 
another is the action of the stellar winds from massive stars and the expansion 
of ionized HII bubbles that can transfer momentum, kinetic energy and heat to 
a nearby gas cloud.  

Understanding the star formation history of a cloud requires an indicator of age and 
evolutionary status. The ratio of stars without disks to stars still bearing a disk 
can be used to assess the degree of evolution of the process of star formation in a gas
cloud \citep{Strom1995,Haisch2001,Gutermuth09}, 
but assessing a complete census of both classes of objects is difficult. 
For regions distant more than 1~kpc, proper motion studies have insufficient sensitivity, 
photometric selection is strongly contaminated by field stars, and spectroscopic
selection requires an enormous effort to study hundreds or
thousands of stars -- even when using new multi-fiber facilities.
Infrared (IR) studies are helpful (when crowding is not extreme), but
they are sensitive only to stars with circumstellar material, bright in IR, but insensitive to objects 
without disks, i.e., Class III Pre Main Sequence (PMS) stars  \citep{Lada1992}. 
However, PMS stars are up to 10$^3$ brighter than the Sun in X-rays, and this fact can be
used to effectively detect young disk-less members of star forming regions and in young 
associations.

The complex of Rho Ophiuchi Dark Cloud is a patchy, multi-core cloud that spans a few degrees in the sky
near the eponymous system \rhoOph. 
{ A range of distances to the cloud are reported in literature \citep{Wilking08}, 
spanning the range 120 pc to 145 pc; here we will use the distance of 120 pc, obtained 
from VLBI measurements by \citet{Loinard08} and valid for the northern part of the 
cloud where \rhoOph\ is located.  The uncertainty on the distance can produce a systematic 
error of the X-ray luminosities of $\sim0.16$ dex. }
The main core of the cloud, L1688, contains high density gas and dust and a stellar population 
of about 300 Young Stellar Objects (YSOs, \citealp{Gagne04,Wilking05}). The dense core F of the cloud 
has been extensively studied in infrared and X-rays. 
In X-rays, several \chandra\ observations have targeted L1688 for a total duration
of about 400 ks \citep{Imanishi01,Gagne04}.\xmm\ has also observed L1688 with a 35 ks exposure
\citep{Ozawa05},  and with a large program named DROXO (PI: S. Sciortino, exposure $\sim500$ ks) 
aimed at obtaining a deep observation to characterize the  X-rays properties of the YSOs embedded in 
the L1688 core \citep{Giardino07,Flaccomio2009, Pillitteri2010aa}. 
X-ray surveys have demonstrated the frequent variability of YSOs in L1688 \citep[see][]{Montmerle1983}, 
and the first cases of detected neutral Fe line at 6.4 keV in stellar X-ray spectra, 
due to the interaction of high energy photons 
from the central object  and cold material in the circumstellar disk.

In the present work, we investigate a region north to the Cores F and A of the cloud, 
centered on \rhoOph~A+B (Fig. \ref{rgb}), { where a partially open ring of dust
is visible in mid and far IR images from {\em Spitzer}-MIPS and IRAS
 \citep{Schnee2005}}.  The ring or shell
was likely formed by the stellar winds of B stars at its center. 
Other similar structures are reported in literature. 
With GLIMPSE, \citet{Churchwell2006} detected 322 partial and closed rings. 
They argue that the bubbles are primarily formed by hot young stars in massive star formation 
regions and about 13\% enclose known star clusters.  Only three of the 
bubbles are identified with known SNRs, and no bubbles coincide with known planetary nebulae or W-R stars;
this suggests that B stars can effectively produce such bubbles.
In a study of similar structures in the Perseus cloud, \citet{Arce2011}  
suggested that bubbles like the one in \rhoOph\ are formed by the interaction of spherical or very 
wide angle winds powered by the young stars inside.  Two of the twelve shells observed  in Perseus
are powered by high-mass stars close to the cloud, while the others appear to be powered by low- 
or intermediate-mass stars in the cloud. \citet{Arce2011} argue that winds from stars with a mass 
loss rate of about $10^{-8}$  M$_\odot$ yr$^{-1}$ to $10^{-6}$ M$_\odot$ yr$^{-1}$ 
are required to produce the observed shells. 

A portion of the warm dust ring around \rhoOph\ has been partially observed in X-rays with ROSAT
\citep{Snowden1994,Casanova1995,Martin1998}. 
Due to the low PSPC sensitivity and partial spatial coverage, only a handful of 
X-ray sources are reported in the region. 
\citet{Martin1998} noticed that the ratio of Class III to Class II objects (or WTTSs to CTTSs
in their terminology)  is 10:1 in this region. 
They suggested that this high ratio could be attributed to the action of the stellar winds from 
\rhoOph\ on the nearby disks and not ascribed entirely to an evolutionary effect.

With the present \xmm\ observation we obtain a more complete census of disk-less stars around \rhoOph, 
and we aim to understand the impact of the clearing action that \rhoOph\ has made in the surrounding 
interstellar medium. We will relate the star formation history in this part of the cloud 
to that of L1688, and, with the use of X-rays and IR data, we aim to quantify how many YSOs 
are formed in this region and their evolutionary stage.
The paper is structured as follows: in Sect. 2 we describe the observations and the data
analysis, in Sect. 3 we report the results, in Sect. 4 and 5 we discuss our findings and present our 
conclusions.

\section{Observations and data analysis}
\begin{figure*}
\begin{center}
\includegraphics[height=0.66\textheight]{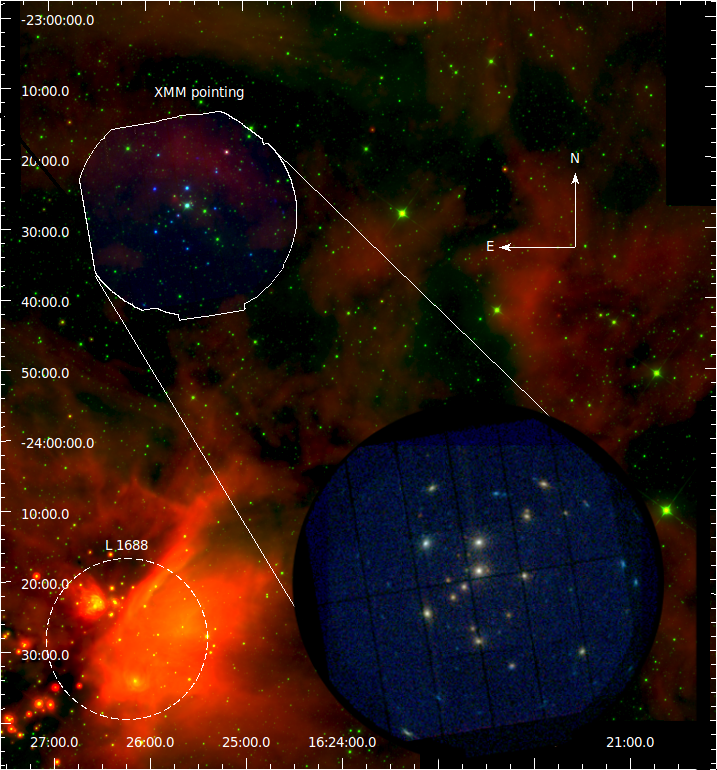}
\end{center}
\caption{\label{rgb} 
Main panel: composite RGB image of WISE images and X-rays. 
Channel bands: blue = EPIC 0.3-8.0 keV; green = 3.4$\mu$m; red = 22 $\mu$m. 
We indicated the \xmm\ pointing and the L1688 region which has been extensively studied.
\rhoOph\ sits at the center of a ring of warm dust that has likely been cleared by its stellar
winds.
Inset image: composite RGB image of EPIC MOS and PN. Channel bands: red = 0.3-1.0 keV; 
green = 1.0-3.0 keV; blue = 3.0-8.0 keV. Different colors of the \xmm\ sources indicate soft (yellow)
or hard (blue) X-ray spectra.
}
\end{figure*}

\begin{table}[t]
\caption{\label{obsx} Log of the \xmm\ observation.}
\begin{center}
\resizebox{0.95\columnwidth}{!}{
\begin{tabular}{c c c c c}\hline \hline
R.A. & Dec. & Filter & Exp. Time & ObsID \\
(J2000) &  (J2000) & & ks & \\
$16^h25^m30.0^s$ & $-23^d28^m00.0^s$ & Thick & 53 & 0720690101 \\ \hline 
\end{tabular}
}
\end{center}
\end{table}
The system of \rhoOph\ A+B ($\alpha = 16^h25^m35.12^s$; $\delta = -23^d26^m49.8^s$) is composed
by a B2IV and a B2V type stars, respectively. 
The system has been observed with \xmm\ on August 29th 2013;
{ Table \ref{obsx} reports the details of the observation.}
Due to the UV brightness of \rhoOph, we used the {\em Thick} filter to
prevent optical photons from triggering spurious events in the EPIC camera, 
which was set as primary instrument, and the OM instruments was not used.

We reduced the ODFs with SAS ver. 13, then we filtered the event tables in the $0.3-8.0$ keV band,
selected only events which triggered at most four adjacent pixels ({\sc PATTERN$<$ 12}) and with
{\sc FLAG = 0}. Toward the end of the exposure when the satellite was approaching 
the orbital perigee, the background level increased. The effect is prominent in PN data, 
while it is negligible in MOS detectors. We filtered out this high background interval 
from the PN exposure,  following the recipe from SAS guide and removing the last 5 ks. 
Fig. \ref{rgb} (inset image) shows a RGB composite image of EPIC data (MOS 1, 2 and PN),
with colors of the sources qualitatively indicating the hardness of their spectra.
{ We chose the following ranges for the RGB channels: $R=0.3-1.0$ keV; $G=1.0-3.0$ keV;
$B=3.0-8.0$ keV. The choice of these energy bands is effective in selecting sources characterized
by thermal/coronal emission and sources with heavily absorbed and harder spectra.}
Sources with soft spectra show yellow/red colors, while sources with harder spectra show blue colors.
{ We used this visual attribute to choose the type of model for fitting the spectra of
sources with enough count statistics.}

For the source detection process, we applied a wavelength convolution algorithm derived from
the analog code developed for ROSAT \citep{Damiani1997a,Damiani1997b} and Chandra. 
The \xmm\ version allows an analysis on the sum of MOS and PN data maximizing the efficiency
of the detection toward the faint sources. We used a threshold of significance equal to 4.6 $\sigma$,
that statistically retain, at most, one spurious source due to background fluctuations.

Spectra and light curves of the sources have been obtained by selecting the events from circular 
regions around the positions of the sources, and choosing their radii in order to avoid 
contamination from nearby sources. Typical extraction radii were $15\arcsec$ to $20\arcsec$. 
For background regions, we followed the prescription of SAS
guide, in particular for PN used regions that are approximately at the same distance
of the source from the read out node, because of the changing response along the chip. 

The spectra were analyzed with XSPEC ver. 12.8 \citep{Arnaud1999} with two types of models:
thermal APEC models (up to three components) plus global absorption for 
coronal sources, and power law spectra plus global absorption for low count statistics spectra 
{ and for sources with a hard spectrum}. The aim was to infer $N_H$ 
absorption, characteristic plasma temperatures and emission measures of the thermal components
for the coronal sources, and power law indices and normalization factors
{ for the sources with hard spectra} or with poor count statistics. 
{ In Table \ref{tabfit} we report the results of the best fit to spectra and the
parameters: $N_H$ absorption, temperatures, normalization factors, power law indices, fluxes in
0.3-8.0 keV band, and $\chi^2$ statistics.} 

\section{Results}
We have detected 89 sources at a significance threshold $>4.6 \sigma$ of the local background.
In Table \ref{xsrc}  we list the coordinates, count rates, off-axis distance and 
statistical significance in units of standard deviation ($\sigma$) of local background fluctuation.
To evaluate the nature of the X-ray sources we use the results of \citet{Getman2011}, 
that estimated the number of contaminants in the Chandra survey of Carina Nebula 
\citep{Townsley2011}. These numbers strongly depends on several parameters including the model 
for galactic stars (foreground and background), galactic sources of different nature 
(e.g., compact objects),  extragalactic sources, distance to the star forming region, 
extinction, and surveyed area.
From scaling the numbers given by Getman et al. by the survey coverage, 
we estimate that about 10 foreground stars and a similar number of background stars should be detected in the \xmm\ FOV, with  
many of these objects having WISE counterparts. In addition,  about 30 extragalactic sources should be detected in X-rays, 
but probably they remain below the sensitivity of WISE.  
{ Out of the 89 X-ray sources, 42 of them are without either a WISE or 2MASS counterpart.
In the following sections we will show that the stars related to the \rhoOph\ cloud 
and in the \xmm\ field of view are characterized by an age of $\sim5-10$ Myr 
and extinction of $A_V\sim3$ mag.   Disk-less stars at $120-145$ pc, 
characterized by such age and extinction values, are easily detectable in X-rays and WISE/2MASS.
The absence of near/mid- IR counterparts implies that, on average, these sources have distances not 
closer than 500 pc, and thus they are not related to the cloud. 
A visual inspection of the X-ray image shows that these are mostly faint sources, 
with detection significance comprised in the range 
$6.6-11.8\sigma$ of local background ($10\%-90\%$ quantiles range). 
Their cumulative spectrum is harder than the average spectrum
of coronal sources belonging to the cloud. The  median of energy of 
the 42 sources with no WISE/2MASS counterpart is $\sim2$ keV, with 90\% quantile above 6 keV,
while the sources likely related to the cloud have a median of $\sim1$ keV, and 90\% quantile
around 2 keV.  
The 42 sources are also spread uniformly across the \xmm\ field. 
We conclude that these 42 sources cannot be related to the cloud, 
and are likely background objects.}
 
The remaining 47 X-ray sources are associated with WISE objects, 
some of them mostly clustered towards the center of the XMM field. 
In section \ref{irclass} we classify the WISE objects with X-ray detection based on 
their mid IR photometry. 
About 50\% of these appear to be YSOs and members of the \rhoOph\ cloud. 

We detected X-ray emission from \rhoOph~A+B (source nr. 50, details in \citealp{Pillitteri2014c}), 
\rhoOph~C (source nr. 62, B5V star with a low mass companion, details in Sect. \ref{rhoophC}) and
{ from  Haro~1-4 (DoAr 16, spectral type K4, source nr. 84) which is  
the only Classical T-Tauri star (CTT) in the \xmm\ field of view.} 
 \rhoOph~D (HD 147888), which is a system with two B3/B4 stars, is undetected, this fact hints that
X-ray emission among B type stars is either peculiar (like in \rhoOph~A+B), or requires a low-mass
 companion (this could be the case for \rhoOph~C discussed in Sect. \ref{rhoophC}, 
see also \citealp{Gagne2011} and references therein).

\subsection{Infrared counterparts to X-ray sources} \label{irclass}
The Dark Cloud of Rho Ophiuchi has been observed with {\em Spitzer} as part of the {\em Cores to Disks} ($C2D$)
program  \citet{Evans2003}, devoted to a deep IR scrutiny of the YSOs in the closest Star Forming Regions 
\citep{Gutermuth09,Gutermuth2011,Kryukova2012}. However, the region
around \rhoOph\ and in the field of view of our \xmm\ observation has not been observed 
with {\em Spitzer} IRAC.
 
\begin{figure*}
\begin{center}
\resizebox{0.8\textwidth}{!}{\includegraphics{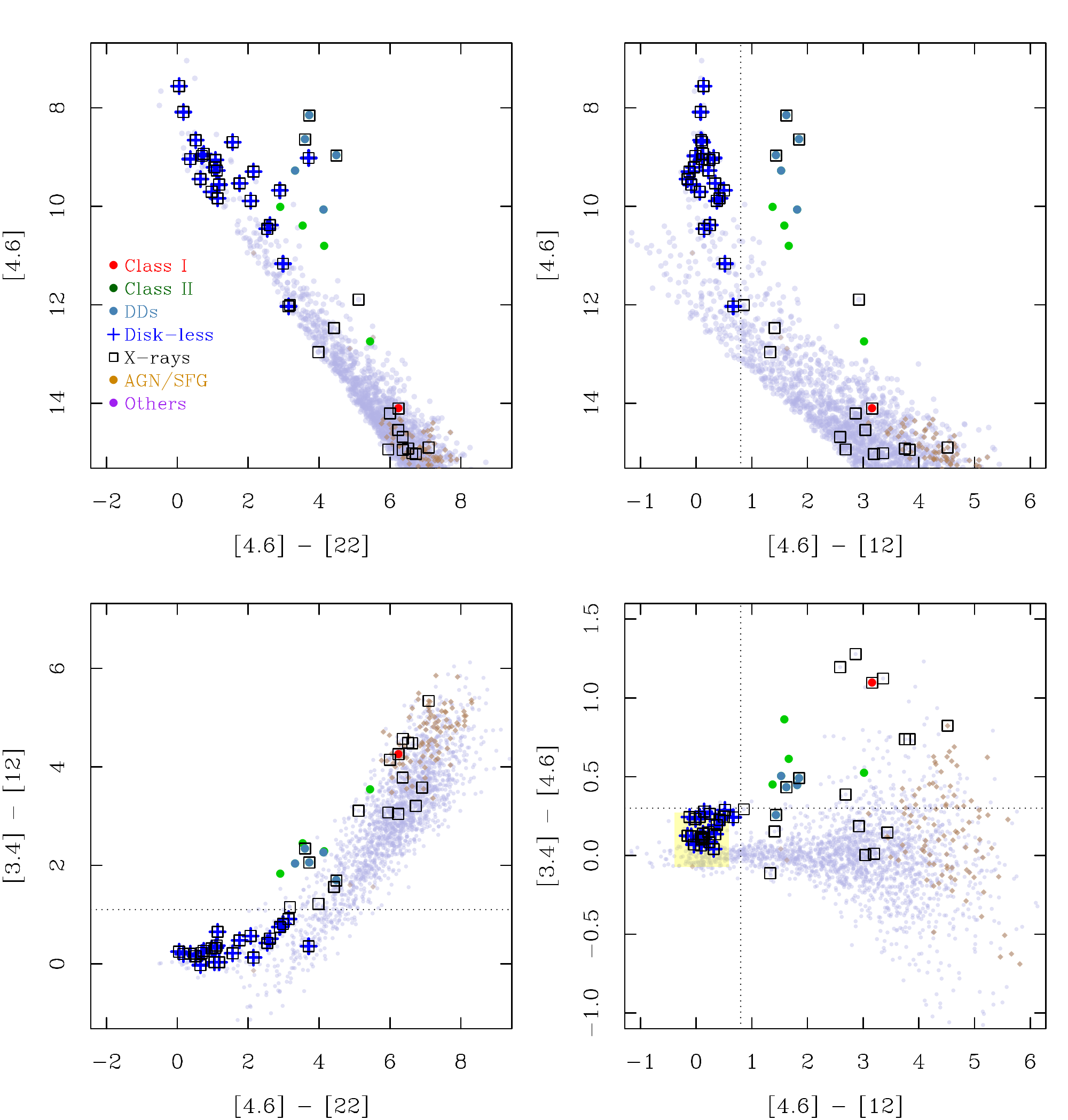}}
\end{center}
\caption{\label{wiseplots} WISE color-magnitude diagrams (top panels) and color-color diagrams (bottom panels) 
for WISE objects within the field of view of \xmm\ ($\sim16\arcmin$ from \rhoOph). 
Symbols are described in the legend in top left panel. { In particular, we mark with 
squares the objects with X-ray emission, and with crosses the objects
with colors typical of normal photospheres (thus not classified as Class I, II or DD objects) that have  
X-ray detection  and $[4.6]<12$ mag. Their number are 22 and are associated with Class III stars.
{ Cuts in magnitudes and colors are marked with dotted lines, these correspond to: $[4.6]-[12]= 0.8$ mag,
$[3.4]-[12] =1.1$ mag, and $[3.4]-[4.6]=0.3$ mag. } }
Very few Class I and Class II YSOs compared to Class III stars are found in the FOV of \xmm\ 
(cf. Fig. \ref{wisel1688} in Appendix). }
\end{figure*}

In order to classify the objects with infrared excesses, we used IR photometry from {\em All WISE} 
catalog \citep{Wright2010}. 
We adopted the scheme of classification described by \citet{Koenig2014},
which uses cuts in WISE colors and magnitudes to classify the YSOs in Class I, Class II and Transition Disks
Objects (also named Debris Disks, hereafter DDs). 
The classification can distinguish YSOs from background AGNs or star-forming galaxies, 
which IR colors can be  similar to YSOs but AGNs have fainter magnitudes than YSOs at the distance of Rho Oph. 
Figure \ref{wiseplots} shows different color-magnitude and color-color diagrams obtained from {\em All WISE} 
photometry of objects in the \xmm\ field. 
Later in Fig. \ref{wisel1688} we show similar plots for the objects in L1688 core. 
In Sect. \ref{discussion} we discuss the difference of
objects with disks and without disks in L1688 and around \rhoOph\ and the implications for the ages
of YSOs and star formation history in the cloud. 

At $9.5\arcmin$ from \rhoOph, Haro~1-4 is classified as a Herbig Ae/Be star with strong H$\alpha$ emission \citep{Maheswar2003}. 
The  scheme of IR classification we adopted fails to assign it to  the Class II / disk stars sample. 
The reason is that the photometry in 3.4$\mu$m band is not reliable as required by the scheme and thus 
Haro~1-4 is formally unclassified.

In Fig. \ref{wiseplots}  we denoted the X-ray sources with WISE counterparts with open squares. 
Among these objects, there are 22 X-ray sources relatively bright in WISE ([4.6]$<12$). 
These objects tend to cluster around (0,0) in the $[3.4]-[4.6]$ vs. $[4.6]-[12]$ color-color diagram 
(marked with crosses in the WISE diagrams). 
We use X-ray detection as a criterion to classify these objects as young stars with X-ray 
bright coronae and normal photospheres. These stars are the disk-less members of the ``Rho Ophiuchi cluster''. 
{ Cuts in magnitudes and colors used to denote the loci of Class III stars are shown in Fig. 
\ref{wiseplots}, these cuts correspond to: $[4.6]-[12] \le 0.8$ mag, $[3.4]-[12] \le1.1$ mag, 
and $[3.4]-[4.6]\le0.3$ mag.}

In the same color-color diagram, the DDs, Class II YSOs and Class I YSOs form a sequence 
distinct from the bulk of background star-forming galaxies and AGNs, 
which are also faint ($[4.6]> 12$) in the  other WISE color-magnitudes diagrams.
The value $[4.6] \sim 12$ corresponds to a spectral type of M6 and $T_\mathrm{eff}\sim3000$ K 
at the distance of \rhoOph\ \citep{Siess2000}.
The value $[4.6] < 12$ allows us to distinguish between YSOs at the distance of \rhoOph\ (120 pc) and background objects. 
This is analogous to the cut in IRAC [4.5] magnitude we adopted in \citet{Pillitteri2013} 
to filter out faint background objects in Orion~A at d$\sim415$ pc with $[4.5] > 14$.
{ This selection is valid only around \rhoOph, where extinction is much lower than, eg.g., in L1688
core.} 
Moreover, the few Class II and Class I YSOs fainter than  $[4.6] = 12$ are likely not members 
of the \rhoOph\ cloud. 
In particular, this is true for source nr. 81, which has IR counterpart classified as a faint Class I YSO. 
However, it has a hard X-ray spectrum which is best fit by a power law, its WISE [4.6] magnitude is 14.11 and thus it 
looks too faint to be a cluster member. For these reasons we excluded source nr. 81 from the list of cluster members.

{ Only 3 objects are questionable in the range $12<[4.6]<13$ and none of them is classified
as Class II or Class I.
In Fig. \ref{wiseplots}, bottom right panel, objects  with $[3.4]-[4.6] > 0.4$ have 
also $[4.6]-[12] > 2.5$. As shown in top right panel, all objects 
with $[4.6]-[12] > 2.5$ have $[4.6]\ge 14$ but one, which has $[4.6]-[12]\sim3$ mag and $[4.6]\sim12$ mag. 
However, this object is not classified as a Class II YSO and still remains too red to be a Class III star 
or a DD objects as well.}

\begin{figure*}
\begin{center}
\resizebox{0.8\textwidth}{!}{
		\includegraphics{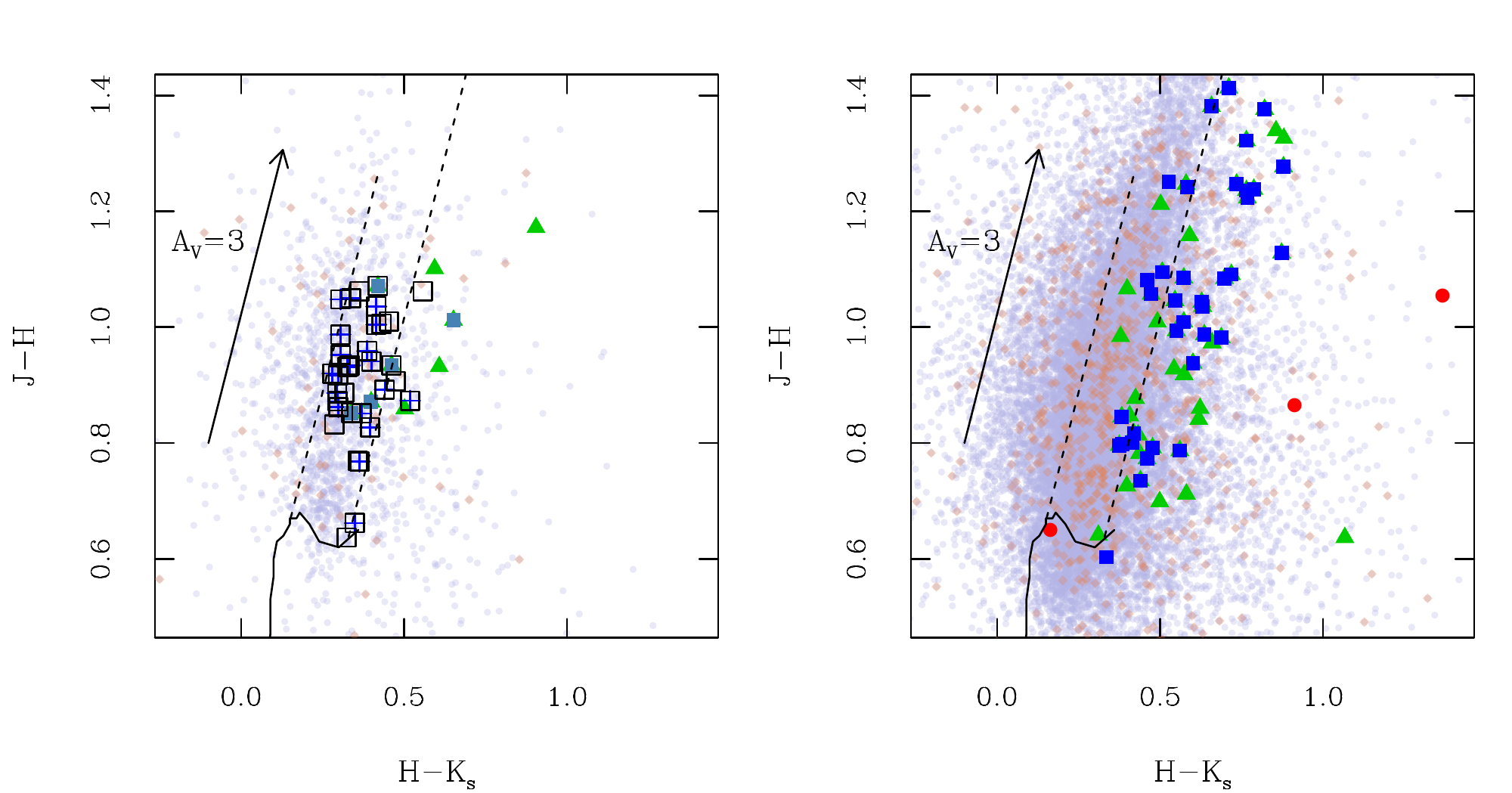}\newline
}
\resizebox{0.8\textwidth}{!}{
		\includegraphics{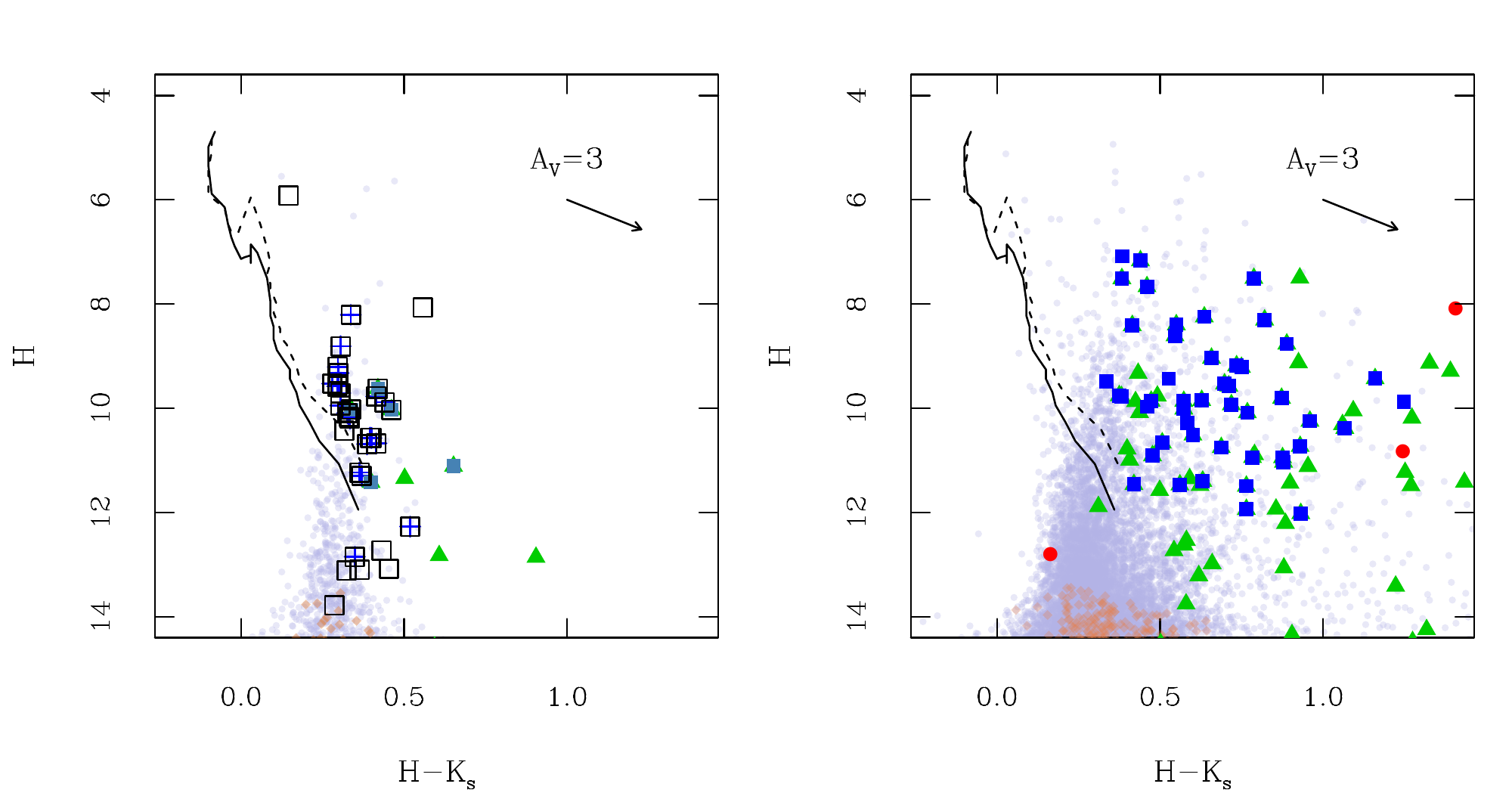}
}
\end{center}
\caption{\label{2masscmd} Top panels: 2MASS $J-H$ vs. $H-K_s$ for objects around \rhoOph\ (left panel), 
and for L1688  (right panel). Bottom panels: $H$ vs. $H - K_s$ for the same two regions.
Isochrones at 10 Myr (solid line) and 5 Myr (dashed line) \citep{Siess2000}  at the distance of 
\rhoOph\ (120 pc) are traced in both panels. Symbols are as in Fig. \ref{wiseplots}.  
Reddening vectors corresponding to $A_V=3$ are indicated. $A_V=3$ is close to the extinction value derived from
gas absorption $N_H$ around \rhoOph, and this is about ten times lower than in L1688.
{ An age comprised between 5 and 10 Myr for most of objects is also inferred from the best fit 
of $L_\mathrm{bol}$ and $T_\mathrm{eff}$ (Sect. \ref{sedsfit}).}  }
\end{figure*}

In Fig. \ref{2masscmd} we show the color-color and the color-magnitude diagrams from 2MASS photometry 
for objects around \rhoOph\  and, for comparison, for objects in L1688. 
The color-color diagram  is useful to classify different IR classes of YSOs \citep{Lada1992} when 
using only $J, H,$ and $K_s$ bands.
The isochrones at 5 and 10 Myr \citep{Siess2000} are traced, as well the reddening vector
corresponding to $A_V = 3$ mag. This value of visual extinction corresponds to an $E(B-V)\sim 1$ mag, 
which is  derived from dust extinction probed by IRAS 100 $\mu$m images  \citep{Schlafly2011}.  
Objects near \rhoOph\ and with X-ray detection are lightly extincted. 
A few objects suffer $A_V\le 2$ mag and all of them fall within the reddening strip of  
absorption/extinction solely expected to intervening material along the line of sight.  
These objects correspond to the Class III sample with no IR excesses. 
By comparison, Class II objects are redder in $H-K_s$ color due to the extra 
emission from their inner disks, and thus they fall on the red side of the reddening strip.

In summary, we obtained a classification of the WISE counterparts of the X-ray sources and 
identified 22 Class III stars and 3 DDs emitting X-rays, these are identified as young members 
of the ``Rho Oph cluster''. 
To these stars we add also \rhoOph\ itself, \rhoOph~C, \rhoOph~D (undetected in X-rays) and Haro~1-4,
and these stars constitute the most massive members of the cluster. 
It is unlikely that we have missed any Class III stars faint in X-rays 
because the limit sensitivity of  the exposure corresponds to $\log L_X\sim27.7$ dex at the distance of 
120~pc, while the typical luminosity of Class III stars is in the range $28.5<\log L_X <30$ dex. 
{ The paucity of Class I/II YSOs in the \xmm\ field points to a real 
difference of  age and evolutionary stage of YSOs in this region with respect
to L1688. This will be discussed later more in details with the support of the result of fitting to 
isochrones of temperatures and bolometric luminosities.} 

\subsection{SEDs, effective temperatures, bolometric luminosities, masses and ages}
\label{sedsfit}
\begin{figure*}
\begin{center}
\resizebox{0.9\textwidth}{!}{\includegraphics{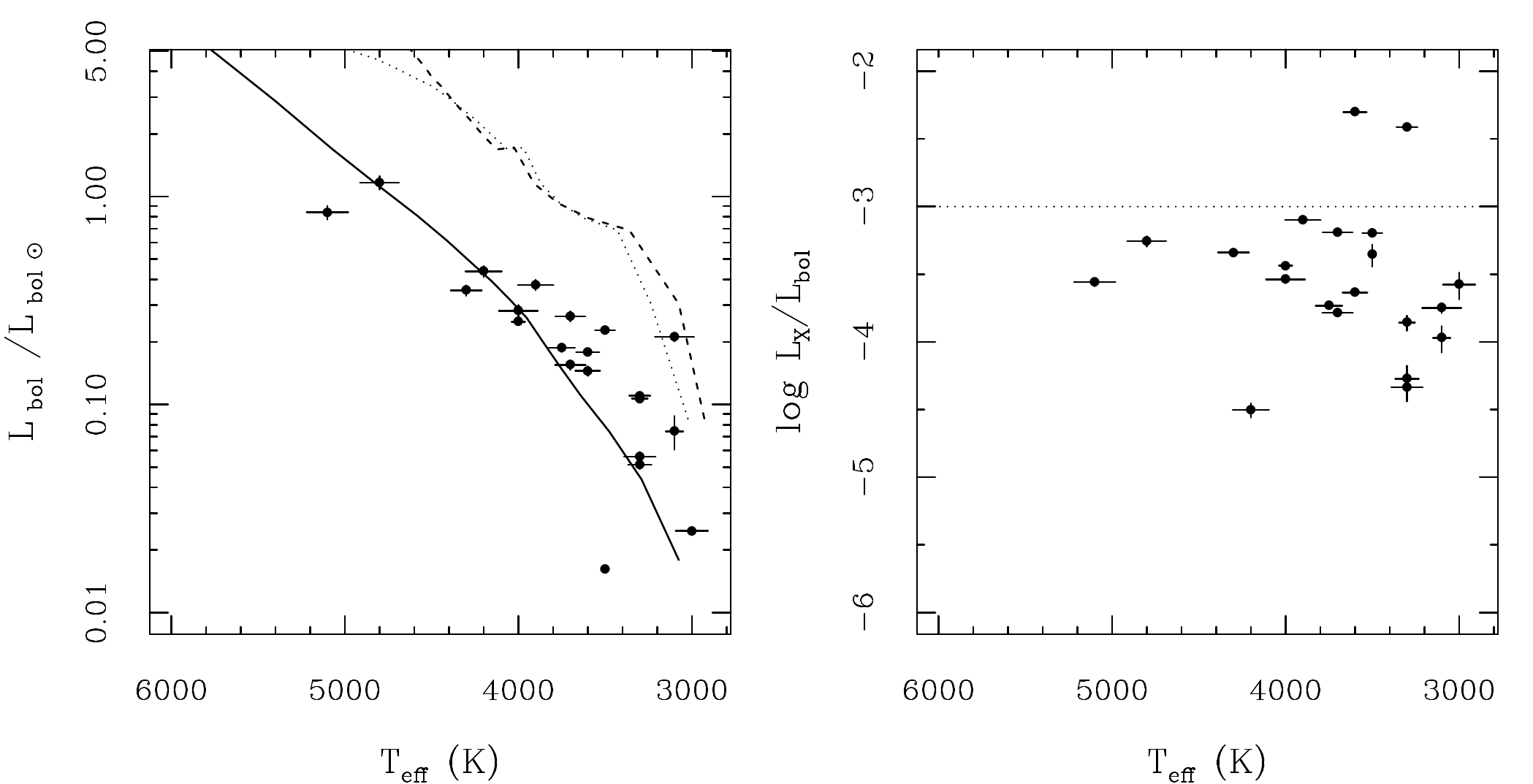}}
\end{center}
\caption{\label{hrd} Left: { $L_{bol}$ vs. $T_\mathrm{eff}$ for Class III stars and DDs in \rhoOph\ cluster 
with spectral fit to SEDs.} Isochrones from \citet{Siess2000} at 1 Myr (dashed line), 5 Myr 
(dotted line) and 10 Myr (solid line) are traced.   
{ Most of the points are comprised betweeen 5 and 10 Myr isochrones.}   
Right panel: $L_X /L{bol}$ ratio vs. $T_\mathrm{eff}$ for YSOs in the \rhoOph\ cluster. 
Most of the objects lie below the saturation limit of  $\log L_X /L{bol}-3$ (horizontal line),
with the only exception of sources 1 and 61 which are found above it. The cause could be a   
systematic uncertainty by $\sim0.6$ dex in $L_\mathrm{bol}$ or $L_X$. 
The points show an overall scatter of about 0.5 dex, and a slight decrease of the  $L_X /L{bol}$ ratio 
is visible for $T_\mathrm{eff}<3500$. }
\end{figure*}

We used VOSA\footnote{\url{http://svo2.cab.inta-csic.es/theory/vosa4/index.php}} web service \citep{Bayo2008} 
to retrieve and fit the Spectral Energy Distributions (SEDs) of the counterparts to X-ray sources.
We searched the available photometry of objects within a radius of 6$\arcsec$ from the X-ray positions. 
{ We restricted our analysis to the X-ray sources that are likely associated to the cloud, and thus
the Class III stars, the DDs, \rhoOph~A+B and \rhoOph~C. We assumed a distance of $120\pm10$ pc and
a visual extinction $A_V=3\pm1$ mag.}
The SEDs were fit with several models: black-body, BT-COND \citep{Allard2012},  BT-DUSTY  \citep{Allard2012},
BT-NEXTGEN-GNS93  \citep{Allard2012},  COND00 \citep{Allard2001,Baraffe2003}, DUSTY00 \citep{Allard2001,Chabrier2000}. The models depend on a set of parameters, including, e.g., $T_\mathrm{eff}$ for the black-body, 
and $\log g$, metallicity, alpha elements enhancement factor for the various versions of BT/NEXTGEN models. 
The number of points in each SED, i.e., the amount of photometry available for each object and its quality,
is crucial for a robust estimate of the best fit parameters of the SEDs. 
In particular, for a number of objects the photometry in U, B, and V bands is missing
due to the moderate extinction in the region (of order of $A_V\sim3.8$ mag). 
This has the effect of making the shape of the SED uncertain. Even in the simplest case of 
black-body model, the peak of the SED  and the $T_\mathrm{eff}$ have errors associated with these color 
uncertainties. 
Also, the reliability of $L_\mathrm{bol}$ is a function of the assumed distance and of the extinction. 
Masses and ages depend on the fit to the isochrones (done through VOSA), and depend on the estimates
of $T_\mathrm{eff}$, $L_\mathrm{bol}$ and the set of isochrones associated with the best fit model. 
Best fit parameters from the VOSA procedure are given in Table \ref{vosafit}. 
Fig. \ref{hrd} shows $L_\mathrm{bol}$ and the ratio $L_X / L_{bol}$ vs. $T_\mathrm{eff}$, 
where  $L_\mathrm{bol}$ and  $T_\mathrm{eff}$ have been obtained from the SED best fits. 
We observe that most of the members of the cluster are found between the isochrones of 5 and 10 Myr, 
with effective temperatures comprised in 3000-5500 K. 
The ages are in the range $2-90$ Myr, with a $10\%-90\%$ range of $3-18$ Myr, these ages appear  
markedly older than the age of objects embedded in L1688. We will discuss the implications of the 
age difference in Sect. \ref{discussion}
The masses inferred from the fit to isochrones are in the range $0.1-1.2 M_\odot$, 
with a median of $0.45 M_\odot$.

\subsection{X-ray spectral properties}
For 27 bright X-ray sources with more than 500 counts in the combined EPIC image 
we performed a best fit modeling of their spectra with XSPEC (v 12.8).
We adopted a combination of APEC thermal models plus absorption or an absorbed power law. 
The choice of the model has been driven by the source count statistics and
the colors in the RGB image (see Fig. \ref{rgb}). 
In general for low count statistics faint sources with apparent hard X-ray spectrum we adopted
a power law model, while for sources with softer spectra (hinting a spectrum from a stellar corona), 
we adopted a thermal model. The results are listed in Table \ref{tabfit}.

\begin{figure}
\begin{center}
\resizebox{\columnwidth}{!}{\includegraphics{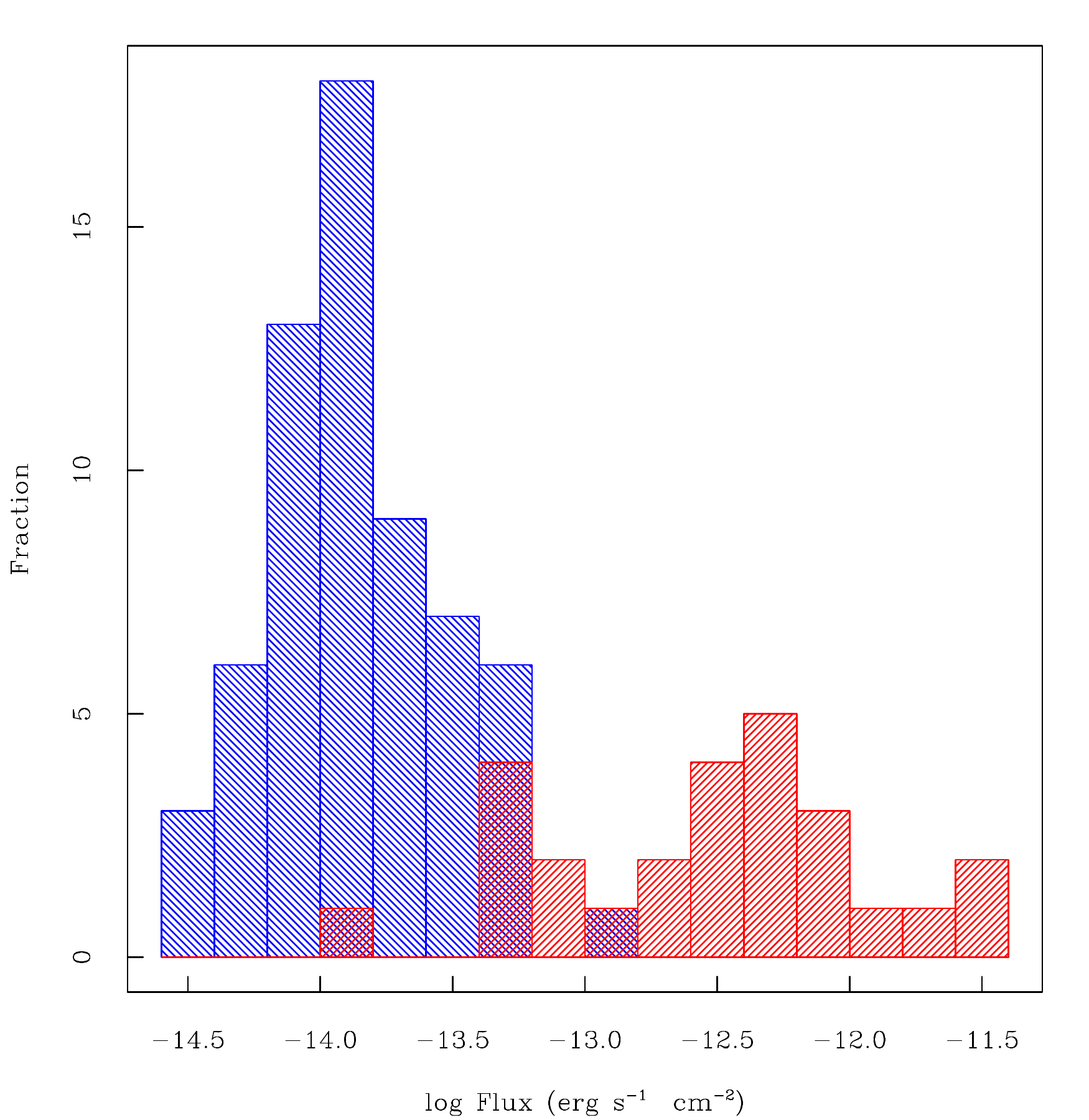} }
\end{center}
\caption{\label{fxhist} Histogram of fluxes obtained from PIMMS (blue bars) and from spectral 
best fit (red bars). 
}
\end{figure} 

In star forming regions it has been observed that the ratio between gas absorption, probed by $N_H$, 
and dust extinction probed by $A_V$ or $A_K$, is not constant. 
The difference could be ascribed to different grain size and flux ablation of grain surfaces 
in those regions with strong ambient UV flux from massive stars.
To compare $N_H$ and $A_V$ we used a map  of  $E(B-V)$ (resolution $\sim1.5\arcmin$)  
based on IRAS and COBE/DIRBE data\footnote{\url{irsa.ipac.caltech.edu/applications/DUST/}} 
\citep{Schlafly2011} to derive $A_V$ by assuming an extinction law with $R_V = 3.1$. 
The average $E(B-V)$ is about 1.02 in the FOV of \xmm,
$N_H$ values are found in a narrow range of values ($20.7 \le \log N_{\mathrm H} (\mathrm{cm}^{-2})\le 21.9$). 
$N_H$ absorption appears is quite uniform across the FOV of \xmm and sensibly lower, by a factor 5 to 10, 
than the values observed in the core F / L1688.

From the extinction map we derive an average $A_V$ of 3.8 mag (10\%-90\% range: $3.6-4.1$ mag) and a
median $N_H/ A_V$ ratio of $1.04\times 10^{21}$ cm$^{-2}$ mag$^{-1}$.
The average $N_H/ A_V$ ratio near \rhoOph\ is lower than in Orion or the typical ISM value,
and slightly higher than the values of $N_H/A_V$ found in other low mass star 
forming regions like NGC 1333, Serpens and L1641 \citep{Winston2007,Winston2010,Pillitteri2013}.
This evidence could be related to different amounts of ambient UV fluxes. 
High UV fluxes can ablate the surface of dust grains and hamper their growth modifying the overall 
properties of dust scattering. Another way to change the ratio is to selectively 
deplete the gas in favor of dust and this could be the mechanism at work in presence of stellar 
winds from B stars.

It is unlikely that the selection of bright X-ray sources for the spectral analysis could have 
introduced a bias toward sources at the surface of the cloud and thus with lower absorption. 
Hot young coronae at the distance of Rho Oph cloud and with plasma temperatures at $1-2$ keV, 
even when highly absorbed,  are still bright enough above 1 keV, to have sufficient 
count statistics in a 50 ks \xmm\ exposure. In this respect, we consider complete the bright sample and 
not biased toward sources with low absorption. 
The source with the lowest absorption is Src 79, it has no counterpart in 2MASS and its X-ray
spectrum is best fitted with a power law of index $\alpha = 1.30\pm0.13$. 

The range of plasma temperatures obtained from  best fit to { models with only one thermal 
APEC component} is fairly limited: $kT = 0.24-1.7$ keV ($10\%-90\%$), with mean of 1.1 keV. 
The best fits of models with two components give a mean second temperature of 1.7 keV, 
and those with 3 components give a mean third component of 3.2 keV.
These values are in agreement with the temperatures of plasma in young coronae of other star forming regions.

\subsection{X-ray luminosities}
{ For the remaining 62 faint sources} we derived a median energy from the distribution of source 
events and used PIMMS software to derive a flux assuming a thermal APEC model with only one component. 
For $N_H$ we used a weighted mean of the $N_H$ values from the best fit to spectra of the bright sources, 
with the weight being the inverse of the spatial distance to the positions of sources in the bright sample. 
This approach is almost equivalent to considering an unique value taken from the average $N_H$, 
given that the absorption is quite uniform across the FOV of \xmm\ and small
variations are expected from source to source. 
Fig. \ref{fxhist} shows the histogram of fluxes derived from PIMMS (green) and from XSPEC (red) for all X-ray sources.
The range is in $-14.56<\log f_X <-11.5$. { Errors in fluxes from PIMMS can be derived from the
uncertainties in count rates; systematic errors due to a different plasma temperature can amount to about 
10\% for temperatures in $0.5-2$ keV. The relative errors of PIMMS fluxes have a median of $\sim17\%$,
 and a $25\%-75\%$ quantile range of $8\%-24\%$.  
For XSPEC derived fluxes, these have relative errors with a median of 3.5\% and a $25\%-75\%$ quantile range of
$2\%-8\%$.
} 
At the distance of \rhoOph\ the minimum detected flux corresponds to $\log L_X \sim 27.7$.
Assuming a complete detection above $\log L_X\sim 28$ and a saturated $L_X/L_{bol}$ ratio, this would imply a detection of
YSOs with $L_{bol} \sim 10^{32}$ erg s$^{-1} \sim 0.1 L_{bol \odot}$.
This is approximately the luminosity of stars of M3-M4 spectral type at 5 Myr \citep{Siess2000}, which represent 
the low mass limit of the X-ray detections, although we presume that we have completeness only at late K- / early M- types.

The ratio $L_X / L_{bol}$ (Fig. \ref{hrd}, right panel) shows a dynamical range
of  three orders of magnitude. The cluster members are distributed in a sequence of values that goes from 
\rhoOph~A+B and \rhoOph~C ($L_X / L_{bol} = 2\times10^{-6}$ and $4\times10^{-5}$, respectively) 
to cooler objects which ratios peak at around $10^{-3}$, 
a value found among the most active and young coronae \citep{Caillault1985,Micela1985,Stauffer94}. 
The origin of this value is debated but it could be a limit to the efficiency in the production of 
X-rays in dynamo driven coronae of active stars \citep{Charbonneau1992} or a breaking of magnetic loops
in fast rotating stars \citep{Jardine99}. 

\begin{figure}
\begin{center}
\resizebox{\columnwidth}{!}{\includegraphics{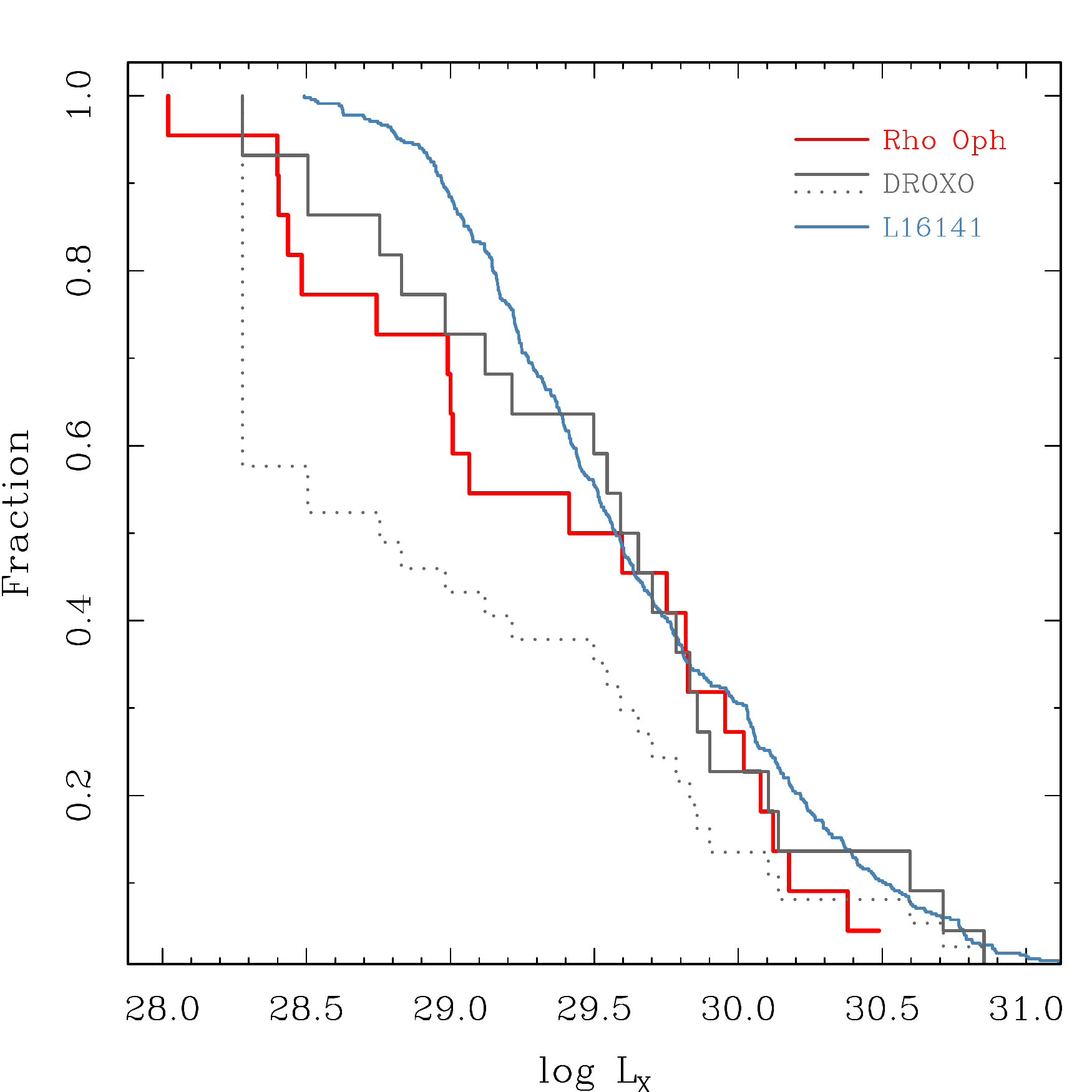} }
\end{center}
\caption{\label{xlf} X-ray luminosity function of Class III stars (comprised 3 DDs). 
For comparison we added the curves of Class III stars in L1688 from DROXO, both corrected 
(solid gray line) and total sample (dotted gray line, see \citealp{Pillitteri2010aa}), 
and L1641 (solid light blue line, \citealp{Pillitteri2013}). 
Class III stars around \rhoOph\ shows less luminosities and a larger range of values.}
\end{figure}

Fig. \ref{xlf} shows the distribution of X-ray luminosities of Class III stars earlier than M5 
around \rhoOph\ and the analog curves of stars in L1688 (from DROXO; \citealp{Pillitteri2010aa}), 
and from L1641 \citep{Pillitteri2013}. 
For L1688 we report the total and a {\em corrected} distribution after removing some suspicious members
from the list of \citet{Bontemps01}, mostly upper limits to luminosity. 
Given the deep sensitivity of DROXO, it is quite unlikely that 
Class III stars remain undetected and thus the corrected distribution is more reliable than the 
distribution of the full sample. 
{ The three distributions tend to be similar, with  
a slight under luminosity of the group of stars around \rhoOph\  for values below $\log L_X \sim29.5$. 
In ONC the X-ray luminosities of  non accreting stars in the range of mass $0.3-1.0$ M$_\sun$ have values around
$\log L_X \sim 29-29.5$ erg/s \citep{Preibisch05}, and similar values are found in Taurus Molecular Cloud 
\citep{Guedel07}. Another small difference is 
the standard deviation of $\log L_X$ is 0.74 dex for Rho Oph, 0.69 dex in L1688 and 
0.59 dex in L1641. Similar differences are found in Classical versus Weak T-Tauri stars \citep{Telleschi07}.
 However, this effect is minor when comparing entire populations \citep{Kuhn2015,Feigelson2011} and cannot rule out 
the hypothesis of an {\em universal}  X-ray luminosity function for PMS stars as proposed by \citet{Feigelson05}.} 

\subsection{X-ray variability}
Strong X-ray variability is detected in several sources. Light curves of the brightest sources 
are shown in Appendix (Fig. \ref{lcs}). 
Here we detail the most prominent cases. 
\subsubsection{\rhoOph~A+B}
Variable X-ray emission from the \rhoOph~A+B system has been detected, this variability has been 
studied in details and published in a separate paper \citep{Pillitteri2014c}.
Briefly, smooth variability was observed in \rhoOph\ on a time scale of 10 ks.
The flux has a low state in the first 30 ks of observation, a slow rise for about 10 ks, 
and a high state for the rest of the observation, 
with a component of $ kT \sim 3.0$ keV gradually appearing in the
spectrum during the rise. The interpretations is that a hot spot of magnetic origin appears 
on the visible face of the primary star because of the stellar rotation ($\ge300$ km/s). 
We inferred a size of the spot of about half of the stellar radius if the spot is at the equator. 
The magnetic origin of this spot is interesting given so few cases of magnetic B2 stars
known in literature. In this respect, \rhoOph\ is likely the best 
target where to investigate the origin of magnetic fields in massive 
stars due to its proximity.

\begin{table}[!t]
\caption{\label{fit_rhoOphC} Best fit parameters of the spectral analysis of \rhoOph~C.
The models are 2T APEC plus global absorption. The non flaring intervals are approximately
the first 8 ks and the interval after the second flare (see Fig. \ref{lcs}.}
\resizebox{\columnwidth}{!}{
\begin{tabular}{l | r | r | r} \hline\hline
 Intervals: &   Quiescent    	&  Flare 1 & Flare 2 \\\hline
 $N_H$  (cm$^{-2}$)		&  $0.38\pm0.02$   & $0.42\pm0.03$ & $0.41\pm0.03$ \\
 $kT_1$ (keV)			&  $1.01\pm0.02$   & $0.97\pm0.05$ & $1.02\pm0.04$ \\
 $Z/Z_\odot$			&  $0.11\pm0.02$   & $0.35\pm0.11$ & $0.35\pm0.13$ \\
 N$_1 (10^{-4}$ cm$^{-5}$)	&  $16.8\pm2.4$	   &  $6.7\pm2.5$  & $8\pm3$       \\
 $kT_2$  (keV) 			&  $3.7\pm0.6$     & $2.9\pm0.2$   & $3.0\pm0.4$   \\
 N$_2 (10^{-4}$ cm$^{-5}$)   	& $5.1\pm0.9$      & $13.6\pm1$    & $10.4\pm1.1$  \\
 Flux (10$^{-12}$ erg s$^{-1}$ cm$^{-2}$) & 1.96   & 2.7           & 2.43          \\
 $\chi^2$   			&156.7  	   & 134.3         & 65.7          \\
 Degrees of freedom  			& 160 		   & 89		   & 79            \\
\hline 
\end{tabular}
}
\end{table}

\subsubsection{\rhoOph~C}\label{rhoophC}
\rhoOph~C (HD 147932, source nr. 62) is composed by a B5 type star ($V=5.8$ mag) and a low mass companion 
($V=11.7$, \citealp{Mason2001}). Recently, it has been discovered that the primary possesses an intense 
magnetic field  ($>3000$ G, \citealp{Alecian2014}). 
This star has a peculiar optical spectrum with weak He I lines. It is a fast rotator, with v$\sin\,i \sim$ 140 km/s, 
and displays changes in the spectra taken one night apart. These changes can be ascribed to rotational features. 

\rhoOph~C exhibited two bright flares with very similar peak intensities and decay times 
(see curve of Src. 62 of Fig. \ref{lcs} in Appendix). 
Other modulated emission is also recognized.
The decay time of both flares is approximately 4~ks, 
with the quiescent level reached at $\sim$20~ks for the first flare and at $\sim31$ ks 
for the second flare.
Table \ref{fit_rhoOphC} reports the best fit results of modeling the spectra during the flares and during the quiescent
state, which has a best fit with a combination of two absorbed APEC thermal components. 
We have tested to model the flaring spectra with three components, keeping two components fixed to the quiescent best fit model, 
however the fit gave an unrealistically high temperature with a negligible component normalization. 
On the other hand, models with two thermal components gave a good fit for the flare spectra. 
In this case, the hot component of the two flaring spectra was found similar to the hot 
component of the spectrum post-flares, albeit with a normalization larger by one order of magnitude. 

Modeling of magnetically confined plasma in solar coronal loops has been extensively studied 
by \citet{Reale2007} in order to develop diagnostics for stellar flares. 
We can apply this modeling to the flare light curve in order to derive a loop length, by taking into account the 
rise and decay times (1.8 ks and 10 ks, respectively) and the peak plasma temperature derived from the flaring intervals 
in Table \ref{fit_rhoOphC}. Using eq. 12)\ in \citet{Reale2007}, we find a loop semi-length of $\sim4\times10^{10}$ cm for both 
flares. It is unclear which star of the \rhoOph~C system flared. 
Young K type stars have enhanced variability and do flare more often than B late type stars \citep{Pye2015}.
The companion of \rhoOph~C is a late K type star ($V=11.7$) at an age of $\sim 5$ Myr and it is expected to be 
 very active and variable in X-rays. Given these characteristics it is plausible that this star hosted both flares. 
The loop semi-length of both flares would be comparable with the stellar radius. 
Were the flares generated in a magnetized corona of the B7 primary, this would be among the few examples
of flare-like variability detected in a B late type star.  

\subsubsection{Other variable sources}
In addition to \rhoOph~C, we find impulsive variability in sources nr. 10, 38 and 41. 
Sources 17, 20, 31, 71, 83, and 84 show less identifiable variability. In source 84 we see  an
overall decrease of the flux as if it is part of the decay of a long lasting flare. 
Some minor flickering at level of 2 $\sigma$ significance is over imposed to the global trend.

Among the flaring sources 10, 38 and 41, the latter is flaring at the beginning of the observation 
for $t<10$ ks. The subsequent portion of light curve shows some slow decrease after
minor impulses. We have analyzed the spectra of the main flare and of the post-flaring intervals.
The best fit model of the post flare spectrum was given by sum of two APEC thin plasma models absorbed 
by a common $N_H$ column. The two temperatures were $kT_1 = 0.93\pm0.04$ keV, $kT_1 = 2.2\pm0.2$,
the normalization factors were $N_1 = (7.2\pm 1)\times 10^{-5}$ cm$^{-5}$ and  $N_2 = (9.3\pm 1.1)\times 10^{-5}$ cm$^{-5}$;
the absorption was $N_H = (2.0\pm 0.3)\times10^{21}$ cm$^{-2}$. For the flare, we used a 3T APEC model with the first
two components fixed to the post flare best fit model. In this case, the third thermal component had a 
best fit value of $kT = 2.5\pm0.3$ keV and normalization $N_3 = (2,3 \pm0.2)\times 10^{-4}$ cm$^{-5}$, i.e., a factor
2 to 2.8 higher than the thermal components of the post-flare spectrum.
By using the scaling laws of \citet{Reale2007} and \citet{Serio1991} as in the case of \rhoOph~C, we estimate a loop semi-length
of $l\ge 5.8\times10^{10}$ cm or 0.8 $R_\odot$. This is a lower limit given that we did not observe the rise and the 
peak of the flare. The post flare spectrum appears somewhat hot, and that could be over dense plasma
flared loop still cooling down after the fast decay happened during the first 10 ks of observation. 
We speculate that the small secondary impulses seen in the light curve could have been triggered by 
the main flare as observed in the Sun in complex active regions \citep[e.g.][]{Aschw2001}.

\section{Discussion}
\label{discussion}
\begin{figure}
\begin{center}
\resizebox{\columnwidth}{!}{\includegraphics{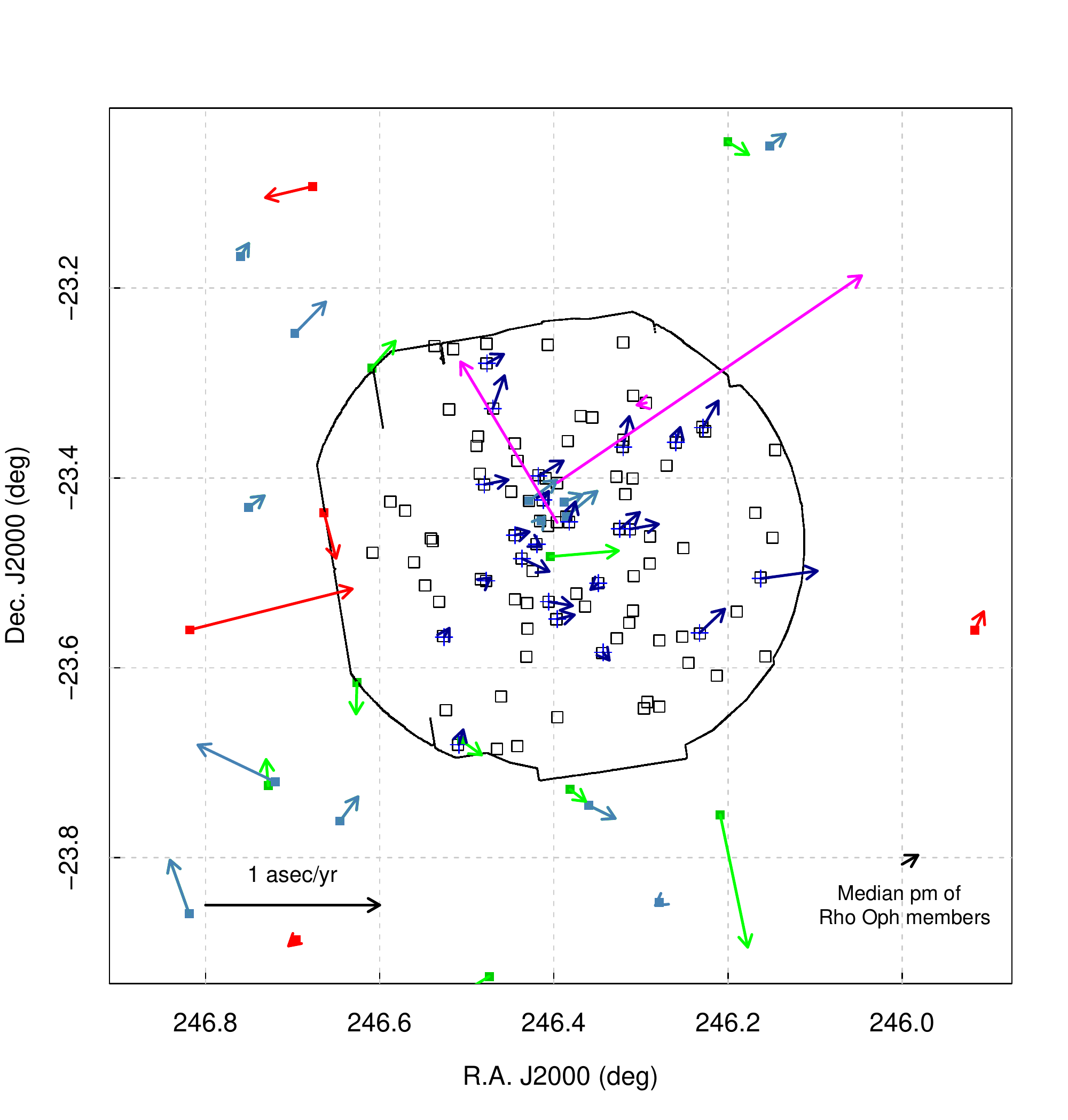} }
\end{center}
\caption{\label{pmotions} Spatial distribution of the X-ray sources (open squares) and YSOs (crosses and 
filled colored symbols) in the \xmm\ FOV with their proper motions (arrows). 
Magenta arrows are for \rhoOph~A+B, \rhoOph~C and Haro 1-4 stars. Red are Class I YSOs, green are Class II YSOs, 
blue are Class III stars. These latter share similar proper motions and form a cluster about 5 Myr old 
formed {\em in situ}, while Class I and Class II YSOs seem to have larger proper motions and have traveled 
across the FOV of \xmm.  The large proper motions of \rhoOph~A+B and \rhoOph~C are likely related 
to their binarity nature. }
\end{figure}

The first issue we discuss is the comparison between \rhoOph\ cluster and YSOs in L1688.
We have shown that there is a cluster 
of about {28 YSOs} surrounding \rhoOph\ and born from the same natal cloud. 
There is evidence of a significant age difference between this cluster and the YSOs in L1688.
The first evidence is pointed out by the comparison of the number of stars with disks  and stars without 
disks in the two regions inferred from WISE and 2MASS photometry and joint IR/X-ray classification.
The ratio of disks/no-disks objects is a statistical probe 
of the evolutionary status of star formation in the cloud and ultimately of the stellar ages \citep[see][]{Haisch2001,Strom1995}.
We count 22 disk-less stars, 3 DD objects and three stars with disks 
({ but two stars with disks are} at the very edge of the \xmm\ FOV, see Fig. \ref{pmotions}) 
near \rhoOph, giving thus a frequency of disks of about 1:8, 
which is similar to what observed with ROSAT  \citep{Martin1998}. 
{ One protostar/Class I object is at the very edge of \xmm\ FOV and it is undetected, 
another Class I object is Haro 1-4 which is detected.} 
In contrast, L1688 core contains a significant number of stars with disks and protostars
embedded in the densest part of the cloud \citep{Gagne04}, and the ratio stars with disks to 
disk-less stars is of order of 1 or more. 
The age inferred from the ratio disks/no-disk objects around \rhoOph\ is about 5 Myr,
while for L1688 embedded YSOs an age $t<1$ Myr is inferred \citep{LR99, Natta2002}.
Other evidence of ages around $5-10$ Myr near \rhoOph\ is given by the 2MASS color-magnitude diagrams
and model isochrones at 5 Myr and 10 Myr, and by the HR diagram obtained from SEDs
modeling and fitting to isochrones (Sect. \ref{sedsfit}).
We can firmly conclude that around \rhoOph\ there is a small cluster a factor of 5 to 10 
older than the age of  YSOs in the L1688 core. 

\citet{Wilking05} analyzed optical spectra from the surface population surrounding L1688, 
deriving spectral types, temperatures, masses and ages for their sample. 
The objects  in \citet{Wilking05} have temperatures in the range $2500-19,000$ K, 
with a mean $T_\mathrm{eff}$  of 3350 K and 90\% quantile at 4380 K. 
This range of $T_\mathrm{eff}$ values  is similar to the range 
we find from SED analysys. There is a similar concurrence among the masses derived by Wilking et al. 
(range: $0.14 \le M/M_\odot \le 0.82$,   mean: $0.22 M_\odot$). 

The ages inferred by \citet{Wilking05} have a peak at $t=2.1$ Myr 
($10\%-\%90$ quantiles range: $0.5 \le t \le 8.2$ Myr). 
As first hypothesized by Wilking et al., the stars clustered around \rhoOph\ were formed in 
a early episode of star formation in the cloud, about $4-5$ Myr ago. 
We also speculate that the densest part cloud was larger in the past and 
that a fraction of the stars have since migrated away from the cloud. What we observe now is the patchy 
``left-over'' cloud of a process of star formation that did not occurred uniformly across the cloud.
The group of disk-less stars we found in X-rays around \rhoOph\ constitute an early product 
of that star forming burst. These stars appear to share similar proper motions 
(Fig. \ref{pmotions}), roughly moving toward north west at 100 mas yr$^{-1}$.  
On the other hand, the few Class I and II YSOs around \rhoOph\ show larger proper motions 
and could have traveled into the region coming from the denser core of L1688.

The Dark Cloud of \rhoOph\ appears to be related to two other nearby star forming regions, 
namely Upper Sco-Cen and Upper Centaurus-Lupus. These also contain YSOs  older than L1688.
\citet{Preibisch1999} suggested that star formation has proceeded starting from Upper 
Centaurus-Lupus 5 Myr ago, then triggered Upper Sco-Cen about 4 Myr ago. 
About 1.5 Myr ago a supernova explosion eventually created the
runaway star $\zeta$ Oph, and the same blast triggered star formation in Rho Oph about 1 Myr ago. 
Our results challenge this scenario. In the north-western edge of L1688 core, star formation has
started earlier than 1 Myr ago, likely 5 Myr ago, and the ``Real Rho Oph cluster" 
appears coeval to Upper Sco-Cen and Upper Centaurus-Lupus. 
The massive stars of the \rhoOph\ systems were formed 
along with other  $\ge25$ solar and sub-solar mass stars north west of L1688. 
Eventually the dust and gas in this part of the cloud has been dispersed by the collective
action of stellar winds from the massive stars in it. 
While it is not ruled out a triggering of star formation due to the progenitor of $\zeta$ Oph, 
this can not have been the solely mechanism of star formation directly associated with Rho Oph.
In our scenario it may be possible that \rhoOph\ had a role in triggering the
star formation also in L1688 at a later phase.

The total X-ray luminosity of the \rhoOph\ cluster amounts to  at least 
$\sim5.6\times10^{30}$ erg s$^{-1}$. 
If this is the minimum rate of emission during a lifetime of  $t=3-5$ Myr,
at least $5\times10^{44}$ erg in X-ray band have been injected into the surrounding cloud. 
If we assume that the stars have a saturated $L_X/L_{bol} \simeq10^{-3}$, 
the total energy deposited into the cloud during the stellar lifetime is up to $\le10^{47}$ erg.  
Chen et al. (in prep.) estimated the total cloud mass (about $\sim3.37\times10^3 M_\odot$), 
energy and velocity of the  expanding bubble around \rhoOph. 
The kinetic energy of the bubble is about $6.7^{45}$ erg, the 
expansion velocity is $\sim 1.3$ km/s and the time to inflate the bubble (which size is about 1.36 pc) 
is $\sim1.2$ Myr. 
The energy entrained in the cloud can be consistent with the energy radiated by the 
young stars within the bubble, with an efficiency factor of $10-15\%$ for converting radiated energy 
into kinetic energy of the bubble. 
However, the age of the bubble appears too short when compared to the 
age of stars in it and this remains an open issue in the connection between stars and natal cloud. 
It is also worth noticing that Chen et al. assumed a constant expansion velocity of the shell throughout 
the lifetime of the shell. While this is generally true for the snowplow phase 
(i.e. when the bubble has expanded beyond the densest part of the cloud), the initial expansion velocity could 
be slower due to the dense material surrounding the young stars when they were born. 
Chen et al. noted that the estimate of the shell formation timescale of 1.2 Myr could be  a lower bound.
We speculate that the efficiency in transferring energy to the cloud
in form of kinetic energy was lower during the very early phases of star formation, 
when the gas of the cloud was denser, this effectively delaying the time of inflation of 
the bubble. 

\section{Conclusions}
We have presented the analysis of an \xmm\ observation of 53 ks centered on the \rhoOph~A+B system. 
We found a total of 89 X-rays sources, 47 of them with WISE counterparts. 
We have classified them according to their WISE colors and magnitudes and used X-ray detection
as a further criterion to classify disk-less stars bright in X-rays. 
In this way we identified a group of 22 young disk-less stars surrounding \rhoOph. These stars 
constitute a small cluster formed  about 5 to 10 Myr ago during a previous event of star formation
just north of L1688 dense core of the cloud. 
Another indicator of age is the almost absence of stars with disks in the 
\xmm\ FOV. In fact, only three Transition Disk / Debris Disk objects are found in it, adding thus evidence
of a more evolved stage of star formation in the region with respect to the core of L1688 with younger YSOs ($t\sim1$ Myr).
This cluster, older than the average age of the well studied
L1688 core of the cloud, challenges the scenario of triggered star formation in Upper Sco-Cen and Rho Oph 
proposed by \citet{Preibisch1999}. In particular, it is evident that star formation has started in the cloud about 5 Myr 
ago simultaneously with Upper Sco-Cen and Upper Centaurus Lupus. Thus YSOs in Rho Oph cloud are not the solely
result of the sweeping action and triggered star formation  due to super novae explosions happened 
in Upper Sco-Cen as proposed by \citet{Preibisch1999}
  
The energy radiated from the stars into the interstellar medium is sufficient to inflate the observed bubble, 
however the time to form the bubble is about five times shorter than the age of the stars in it. 
To solve the issue, we could speculate of a lower efficiency of transformation of the energy radiated
from stars into kinetic energy the cloud in the early phases that delayed the inflation of the bubble. 

Some stars exhibit variability in form of flares or in less classifiable shape. 
The flaring rate appears lower than in the core F of L1688. 
Time resolved spectroscopy of the most intense flares lead to estimate loop lengths
of order of the stellar radius, which are compatible with similar findings in other young active stars.


\begin{acknowledgements}
I.P. is thankful to Dr. Mario Guarcello for the fruitful discussions about the 
lifetime and evaporation of circumstellar disks in PMS stars. 
S.J.W. was supported by NASA contract NAS8-03060.    
This publication makes use of VOSA, developed under the Spanish Virtual 
Observatory project supported from the Spanish MICINN through grant 
AyA2011-24052.
Based on observations obtained with XMM-Newton, an ESA science mission 
with instruments and contributions directly funded by ESA Member States 
and NASA.
\end{acknowledgements}

\onecolumn
\appendix

\section{WISE diagrams of L1688.}
\begin{figure*}
\begin{center}
\resizebox{0.8\textwidth}{!}{\includegraphics{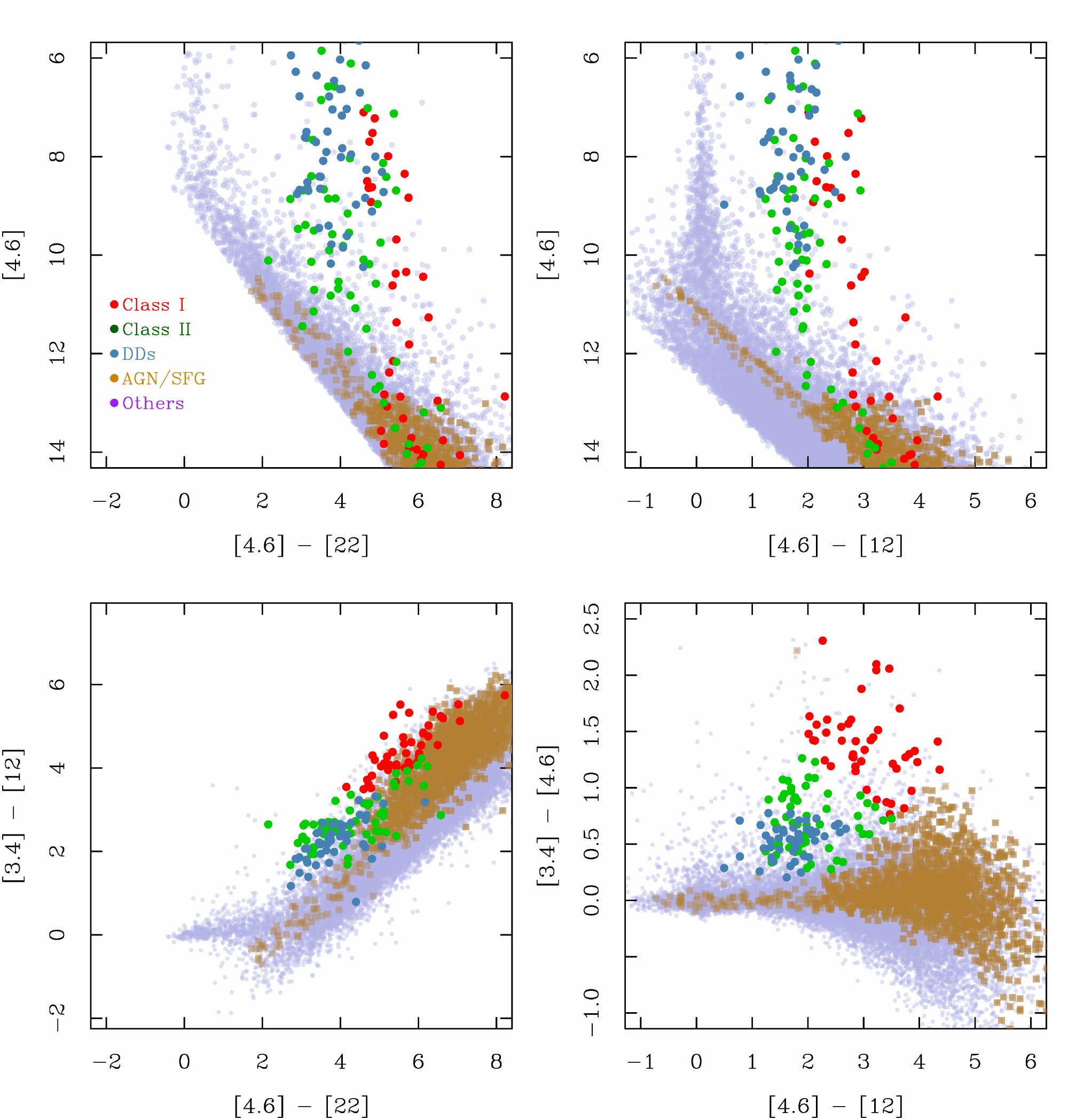}}
\end{center}
\caption{\label{wisel1688} Same as in Fig. \ref{wiseplots} for objects in L1688 within 1 deg from SR 12A. 
Differently from \rhoOph\ region, we find many Class I and Class II YSOs.}
\end{figure*}

\section{List of X-ray sources}
 \begin{table*}[h!]
\begin{center}
 \caption{\label{xsrc} Properties of X-ray sources. Coordinates, positional errors, off-axis angle,
exposure time, count rates, errors on count rates and significance are listed. Flags indicating the
classification based on IR/X-ray emission (see Sect. \ref{irclass}) are also indicated. 
}
\begin{sideways}
\resizebox{!}{0.32\textwidth}{
\begin{tabular}{rrrrrrrrrrr}
  \hline
  \hline
 & RA & Dec & Pos.Err & Offaxis & Exp.time & Count.rate & Err.Rate  & Significance & Flux  \\
 & Deg (J2000) & Deg. (J2000) & arcmin & ks & ct ks$^{-1}$ & ct ks$^{-1}$ &  ct ks$^{-1}$ & $\sigma$ & $10^{-14}$\fxu\ \\ 
  \hline
1 & 246.50999 & -23.68115 & 2.1 & 13.16 & 26.83 & 66.2 & 2.0 & 60.5 & 87.1 \\ 
  2 & 246.39589 & -23.65206 & 3.5 & 9.57 & 55.26 & 1.4 & 0.2 & 8.1 & 2.0 \\ 
  3 & 246.52393 & -23.64472 & 2.7 & 11.81 & 51.61 & 1.1 & 0.2 & 7.4 & 1.6 \\ 
  4 & 246.29637 & -23.64279 & 5.0 & 10.30 & 32.22 & 3.2 & 0.6 & 9.4 & 4.6 \\ 
  5 & 246.27875 & -23.64088 & 4.1 & 10.71 & 56.08 & 0.8 & 0.2 & 5.0 & 1.3 \\ 
  6 & 246.29254 & -23.63603 & 2.4 & 10.05 & 57.04 & 1.3 & 0.2 & 9.2 & 1.9 \\ 
  7 & 246.46064 & -23.63028 & 3.3 & 9.18 & 53.10 & 7.1 & 0.6 & 23.1 & 10.2 \\ 
  8 & 246.21291 & -23.60839 & 4.8 & 11.84 & 66.90 & 1.6 & 0.3 & 8.6 & 2.4 \\ 
  9 & 246.24531 & -23.59472 & 4.5 & 9.92 & 78.63 & 1.6 & 0.2 & 9.3 & 2.4 \\ 
  10 & 246.43154 & -23.58834 & 3.4 & 6.23 & 72.43 & 1.0 & 0.2 & 9.1 & 1.4 \\ 
  11 & 246.15713 & -23.58815 & 3.3 & 13.90 & 26.45 & 3.6 & 0.6 & 11.9 & 5.4 \\ 
  12 & 246.34435 & -23.58452 & 2.1 & 5.99 & 81.44 & 18.6 & 0.6 & 60.3 & 5.7 \\ 
  13 & 246.27888 & -23.57108 & 2.8 & 7.59 & 77.09 & 0.5 & 0.1 & 6.8 & 0.8 \\ 
  14 & 246.32773 & -23.56896 & 4.2 & 5.63 & 85.76 & 0.5 & 0.1 & 4.7 & 0.7 \\ 
  15 & 246.25247 & -23.56733 & 2.6 & 8.66 & 82.91 & 0.6 & 0.1 & 7.0 & 1.0 \\ 
  16 & 246.52719 & -23.56657 & 5.7 & 8.87 & 69.22 & 1.0 & 0.2 & 5.4 & 1.5 \\ 
  17 & 246.23220 & -23.56384 & 1.9 & 9.54 & 84.01 & 34.0 & 0.9 & 91.8 & 38.7 \\ 
  18 & 246.43039 & -23.55880 & 2.2 & 4.61 & 105.31 & 1.0 & 0.1 & 12.6 & 1.4 \\ 
  19 & 246.31344 & -23.55269 & 3.1 & 5.43 & 59.90 & 0.3 & 0.2 & 4.7 & 0.5 \\ 
  20 & 246.39709 & -23.54892 & 1.3 & 3.41 & 86.99 & 41.2 & 1.0 & 118.0 & 50.0 \\ 
  21 & 246.18988 & -23.54087 & 5.8 & 11.24 & 76.65 & 0.8 & 0.2 & 5.8 & 1.2 \\ 
  22 & 246.30907 & -23.53986 & 4.2 & 5.15 & 113.65 & 0.4 & 0.1 & 4.9 & 0.6 \\ 
  23 & 246.36410 & -23.53545 & 3.0 & 2.86 & 127.22 & 1.0 & 0.1 & 12.0 & 1.4 \\ 
  24 & 246.43070 & -23.53176 & 4.1 & 3.34 & 115.46 & 0.4 & 0.1 & 5.5 & 0.6 \\ 
  25 & 246.53167 & -23.53044 & 3.6 & 8.25 & 58.43 & 0.5 & 0.2 & 4.8 & 0.9 \\ 
  26 & 246.40614 & -23.53050 & 1.8 & 2.49 & 123.29 & 5.8 & 0.3 & 41.5 & 6.7 \\ 
  27 & 246.44470 & -23.52781 & 3.3 & 3.79 & 111.57 & 0.9 & 0.1 & 9.4 & 1.3 \\ 
  28 & 246.37404 & -23.52188 & 3.8 & 1.89 & 132.70 & 0.5 & 0.1 & 6.2 & 0.6 \\ 
  29 & 246.54793 & -23.51319 & 4.4 & 8.92 & 70.22 & 1.9 & 0.3 & 10.4 & 2.9 \\ 
  30 & 246.34917 & -23.51089 & 1.9 & 2.37 & 136.95 & 4.2 & 0.2 & 35.8 & 5.8 \\ 
  31 & 246.47790 & -23.50844 & 1.0 & 5.07 & 101.53 & 77.7 & 1.0 & 171.8 & 179.0 \\ 
  32 & 246.48393 & -23.50664 & 1.7 & 5.38 & 47.48 & 0.7 & 0.2 & 8.0 & 1.2 \\ 
  33 & 246.16256 & -23.50487 & 4.7 & 12.39 & 71.97 & 1.6 & 0.3 & 8.3 & 2.7 \\ 
  34 & 246.30824 & -23.50337 & 2.4 & 4.40 & 131.63 & 0.2 & 0.1 & 4.8 & 0.3 \\ 
  35 & 246.42445 & -23.49805 & 3.2 & 2.06 & 126.69 & 0.3 & 0.1 & 4.6 & 0.4 \\ 
  36 & 246.28996 & -23.49029 & 3.4 & 5.36 & 126.81 & 0.7 & 0.1 & 8.6 & 1.0 \\ 
  37 & 246.56058 & -23.48891 & 1.7 & 9.53 & 66.97 & 0.3 & 0.1 & 4.8 & 0.5 \\ 
  38 & 246.43710 & -23.48494 & 1.4 & 2.78 & 123.39 & 15.4 & 0.5 & 76.2 & 52.2 \\ 
  39 & 246.60808 & -23.47868 & 5.9 & 12.17 & 49.96 & 1.3 & 0.3 & 4.8 & 1.9 \\ 
  40 & 246.25135 & -23.47380 & 2.7 & 7.57 & 108.56 & 1.6 & 0.2 & 15.2 & 2.4 \\ 
  41 & 246.41984 & -23.46983 & 1.2 & 2.26 & 131.01 & 24.3 & 0.6 & 105.6 & 33.1 \\ 
  42 & 246.53912 & -23.46646 & 3.9 & 8.50 & 72.53 & 2.7 & 0.3 & 14.7 & 4.1 \\ 
  43 & 246.54110 & -23.46385 & 1.9 & 8.64 & 71.22 & 0.3 & 0.1 & 5.2 & 0.7 \\ 
  44 & 246.14878 & -23.46293 & 1.9 & 13.25 & 28.78 & 28.3 & 1.0 & 45.4 & 42.1 \\ 
  45 & 246.28951 & -23.46172 & 5.6 & 5.70 & 127.71 & 0.6 & 0.1 & 5.4 & 0.9 \\ 
\hline
\end{tabular}
}
\resizebox{!}{0.32\textwidth}{
\begin{tabular}{rrrrrrrrrrr}
  \hline
  \hline
 & RA & Dec & Pos.Err & Offaxis & Exp.time & Count.rate & Err.Rate  & Significance & Flux \\
 & Deg (J2000) & Deg. (J2000) & arcmin & ks & ct ks$^{-1}$ & ct ks$^{-1}$ &  ct ks$^{-1}$ & $\sigma$  & $10^{-14}$\fxu\  \\ 
  \hline
  46 & 246.44489 & -23.46047 & 1.8 & 3.71 & 117.72 & 6.0 & 0.3 & 40.2 & 60.6 \\ 
  47 & 246.31255 & -23.45428 & 3.0 & 4.72 & 134.83 & 0.4 & 0.1 & 5.8 & 0.6 \\ 
  48 & 246.32470 & -23.45381 & 1.2 & 4.17 & 134.06 & 31.5 & 0.6 & 122.2 & 32.7 \\ 
  49 & 246.40655 & -23.45051 & 1.8 & 2.75 & 131.16 & 0.2 & 0.1 & 4.9 & 0.4 \\ 
  50 & 246.39647 & -23.44674 & 0.8 & 2.81 & 133.25 & 141.8 & 1.0 & 280.0 & 300.0 \\ 
  51 & 246.38253 & -23.44644 & 1.8 & 2.79 & 134.36 & 1.8 & 0.2 & 19.9 & 3.2 \\ 
  52 & 246.41584 & -23.44544 & 2.4 & 3.25 & 99.95 & 3.7 & 0.3 & 26.3 & 9.0 \\ 
  53 & 246.38580 & -23.44071 & 2.5 & 3.13 & 135.86 & 0.8 & 0.1 & 8.7 & 1.4 \\ 
  54 & 246.16879 & -23.43680 & 1.9 & 12.49 & 51.91 & 20.3 & 1.0 & 50.6 & 45.1 \\ 
  55 & 246.57079 & -23.43441 & 3.9 & 10.69 & 60.48 & 1.0 & 0.2 & 6.8 & 1.5 \\ 
  56 & 246.58780 & -23.42495 & 4.8 & 11.76 & 52.22 & 2.1 & 0.3 & 8.7 & 3.2 \\ 
  57 & 246.42853 & -23.42416 & 2.8 & 4.70 & 105.61 & 1.5 & 0.2 & 13.2 & 3.0 \\ 
  58 & 246.41281 & -23.42343 & 2.9 & 4.39 & 116.65 & 1.3 & 0.2 & 11.8 & 5.9 \\ 
  59 & 246.31788 & -23.41718 & 3.1 & 5.93 & 126.73 & 0.3 & 0.1 & 4.9 & 0.5 \\ 
  60 & 246.44906 & -23.41442 & 2.5 & 5.80 & 108.13 & 0.6 & 0.1 & 8.6 & 1.0 \\ 
  61 & 246.47999 & -23.40686 & 0.9 & 7.25 & 87.82 & 133.4 & 2.0 & 212.0 & 139.0 \\ 
  62 & 246.39643 & -23.40545 & 0.8 & 5.27 & 118.52 & 159.6 & 2.0 & 278.4 & 306.0 \\ 
  63 & 246.30951 & -23.40045 & 2.3 & 7.01 & 112.75 & 0.2 & 0.1 & 4.8 & 0.3 \\ 
  64 & 246.40947 & -23.40044 & 3.2 & 5.67 & 116.22 & 0.4 & 0.1 & 4.6 & 0.7 \\ 
  65 & 246.32809 & -23.39869 & 4.9 & 6.52 & 118.89 & 0.8 & 0.1 & 7.3 & 1.4 \\ 
  66 & 246.41772 & -23.39724 & 3.6 & 5.97 & 107.07 & 0.8 & 0.1 & 6.9 & 1.6 \\ 
  67 & 246.48503 & -23.39551 & 3.3 & 7.94 & 66.60 & 1.7 & 0.3 & 10.2 & 2.6 \\ 
  68 & 246.27048 & -23.38697 & 2.9 & 9.04 & 100.94 & 0.4 & 0.1 & 6.1 & 0.7 \\ 
  69 & 246.44187 & -23.38201 & 3.5 & 7.29 & 76.35 & 1.0 & 0.2 & 7.9 & 1.5 \\ 
  70 & 246.14586 & -23.37060 & 3.1 & 15.18 & 24.49 & 1.6 & 0.4 & 7.5 & 2.4 \\ 
  71 & 246.32095 & -23.36722 & 1.3 & 8.38 & 68.47 & 40.8 & 1.0 & 100.1 & 69.4 \\ 
  72 & 246.48893 & -23.36626 & 3.9 & 9.43 & 60.17 & 3.1 & 0.3 & 13.5 & 5.9 \\ 
  73 & 246.44489 & -23.36354 & 3.1 & 8.38 & 69.64 & 0.5 & 0.1 & 4.8 & 0.7 \\ 
  74 & 246.25978 & -23.36266 & 1.9 & 10.50 & 74.34 & 8.3 & 0.5 & 37.0 & 15.0 \\ 
  75 & 246.38395 & -23.36119 & 5.5 & 7.90 & 77.97 & 0.9 & 0.2 & 5.5 & 1.4 \\ 
  76 & 246.32029 & -23.35975 & 1.6 & 8.80 & 75.89 & 4.2 & 0.3 & 26.1 & 19.3 \\ 
  77 & 246.48702 & -23.35644 & 3.9 & 9.85 & 58.51 & 1.0 & 0.2 & 6.4 & 1.5 \\ 
  78 & 246.22609 & -23.35087 & 3.2 & 12.30 & 74.42 & 5.4 & 0.4 & 23.1 & 5.7 \\ 
  79 & 246.22945 & -23.34635 & 2.7 & 12.36 & 75.86 & 0.8 & 0.2 & 8.3 & 1.5 \\ 
  80 & 246.35589 & -23.33620 & 2.8 & 9.55 & 67.53 & 2.4 & 0.3 & 13.9 & 4.0 \\ 
  81 & 246.36928 & -23.33467 & 1.7 & 9.54 & 68.23 & 12.7 & 0.6 & 45.7 & 24.5 \\ 
  82 & 246.52030 & -23.32785 & 2.9 & 12.31 & 53.53 & 0.9 & 0.2 & 6.4 & 1.3 \\ 
  83 & 246.46964 & -23.32679 & 1.8 & 10.94 & 54.82 & 58.2 & 1.0 & 94.3 & 76.7 \\ 
  84 & 246.29411 & -23.32092 & 1.2 & 11.52 & 74.59 & 46.4 & 1.0 & 103.9 & 60.1 \\ 
  85 & 246.30893 & -23.31346 & 3.1 & 11.60 & 59.41 & 2.0 & 0.3 & 10.7 & 3.0 \\ 
  86 & 246.47738 & -23.27933 & 2.0 & 13.73 & 42.00 & 1.0 & 0.2 & 7.5 & 1.8 \\ 
  87 & 246.51560 & -23.26431 & 1.7 & 15.42 & 25.88 & 0.9 & 0.3 & 5.8 & 1.4 \\ 
  88 & 246.40692 & -23.25975 & 3.1 & 14.02 & 42.05 & 2.8 & 0.4 & 11.6 & 4.4 \\ 
  89 & 246.32039 & -23.25723 & 3.0 & 14.61 & 42.62 & 0.9 & 0.2 & 5.5 & 1.3 \\ 
\hline
\end{tabular}
}
\end{sideways}
\end{center}
\end{table*}

\section{List of IR counterparts}
\begin{table}
\caption{\label{irphot} WISE and 2MASS photometry of the IR counterparts to X-ray sources. Proper motions from {\em All WISE} catalog are listed.}
\begin{center}
\begin{sideways}
\resizebox{1.3\textwidth}{!}{
\begin{tabular}{rlrrrrrrrrrrrrrrrrrrrrrc}
  \hline
  \hline
Src & Designation & RA (J2000) & Dec (J2000) & [3.4] & e([3.4]) & [4.6] & e([4.6]) & [12] & e([12]) & [22] & e([22]) & $J$ & e($J$) & $H$ & e($H$) & $K_s$ & e($K_s$) & pmRA & pmDec & sep & Flag\\ 
    &             & deg        &  deg        & \multicolumn{14}{|c|}{mag} & \multicolumn{2}{c|}{ mas yr$^{-1}$}     & $\arcsec$ \\
  \hline
  1 & J162602.17-234051.9 & 246.5091 & -23.6811 & 9.92 & 0.02 & 9.68 & 0.02 & 9.18 & 0.04 & 6.78 & 0.10 & 11.51 & 0.02 & 10.57 & 0.02 & 10.17 & 0.02 & -25.00 & 79.00 & 3.00 & III \\ 
   10 & J162543.81-233518.2 & 246.4325 & -23.5884 & 15.33 & 0.06 & 14.94 & 0.10 & 12.25 & 0.53 & 8.99 &  &  &  &  &  &  &  & 296.00 & -1352.00 & 3.40 & \\ 
   12 & J162522.43-233501.1 & 246.3435 & -23.5837 & 9.28 & 0.02 & 9.21 & 0.02 & 9.25 & 0.04 & 8.16 &  & 10.70 & 0.02 & 9.71 & 0.02 & 9.40 & 0.02 & -36.00 & -42.00 & 4.20 & III \\ 
   16 & J162606.27-233403.7 & 246.5261 & -23.5677 & 11.46 & 0.02 & 11.17 & 0.02 & 10.65 & 0.27 & 8.18 & 0.36 & 13.14 & 0.02 & 12.27 & 0.03 & 11.75 & 0.02 & -30.00 & 47.00 & 5.40 & III \\ 
   17 & J162455.82-233347.9 & 246.2326 & -23.5633 & 9.14 & 0.02 & 9.04 & 0.02 & 8.94 & 0.05 & 8.68 &  & 10.77 & 0.02 & 9.77 & 0.03 & 9.36 & 0.02 & -144.00 & 126.00 & 2.30 & III\\ 
   18 & J162542.92-233333.6 & 246.4289 & -23.5593 & 14.55 & 0.04 & 14.54 & 0.07 & 11.51 & 0.33 & 8.31 & 0.31 & 15.91 & 0.08 & 15.15 & 0.08 & 14.79 & 0.11 & 267.00 & 274.00 & 5.40 &  \\ 
   19 & J162515.12-233308.8 & 246.3130 & -23.5525 & 12.08 & 0.02 & 11.90 & 0.02 & 8.97 & 0.06 & 6.78 & 0.10 & 13.73 & 0.03 & 12.73 & 0.02 & 12.30 & 0.03 & -194.00 & 106.00 & 1.60 & \\ 
   20 & J162535.03-233255.2 & 246.3960 & -23.5487 & 8.72 & 0.02 & 8.66 & 0.02 & 8.57 & 0.03 & 8.14 & 0.27 & 10.08 & 0.02 & 9.21 & 0.02 & 8.91 & 0.02 & -98.00 & 22.00 & 3.80 & III \\ 
   26 & J162537.38-233149.3 & 246.4058 & -23.5304 & 9.67 & 0.02 & 9.53 & 0.02 & 9.19 & 0.04 & 7.77 & 0.25 & 11.09 & 0.02 & 10.16 & 0.02 & 9.83 & 0.02 & -136.00 & -21.00 & 1.30 & III \\ 
   30 & J162523.74-233038.7 & 246.3489 & -23.5108 & 9.68 & 0.02 & 9.56 & 0.02 & 9.65 & 0.08 & 8.38 &  & 11.13 & 0.03 & 10.20 & 0.02 & 9.87 & 0.02 & 44.00 & -34.00 & 0.90 & III\\ 
   31 & J162554.61-233029.9 & 246.4776 & -23.5083 & 8.20 & 0.02 & 8.09 & 0.02 & 8.01 & 0.02 & 7.92 & 0.23 & 9.76 & 0.03 & 8.81 & 0.02 & 8.51 & 0.02 & -23.00 & 16.00 & 1.10 & III\\ 
   33 & J162438.72-233017.8 & 246.1614 & -23.5050 & 12.22 & 0.10 & 12.21 & 0.10 & 11.32 & 0.36 & 8.11 &  & 14.16 & 0.04 & 13.10 & 0.04 & 12.74 & 0.04 &  &  & 3.90 & \\ 
   33 & J162438.97-233021.5 & 246.1624 & -23.5060 & 10.08 & 0.03 & 9.89 & 0.02 & 9.52 & 0.07 & 7.82 & 0.33 & 11.71 & 0.02 & 10.68 & 0.02 & 10.26 & 0.02 & -325.00 & 40.00 & 4.10 &  \\ 
   38 & J162544.78-232905.6 & 246.4366 & -23.4849 & 9.21 & 0.02 & 8.98 & 0.02 & 8.99 & 0.04 & 8.29 &  & 10.78 & 0.03 & 9.88 & 0.04 & 9.44 & 0.02 & -156.00 & -68.00 & 1.60 & III \\ 
   40 & J162500.23-232825.1 & 246.2510 & -23.4737 & 12.30 & 0.02 & 12.01 & 0.02 & 11.15 &  & 8.83 &  & 14.09 & 0.03 & 13.08 & 0.03 & 12.62 & 0.03 & 14.00 & -24.00 & 1.40 & \\ 
   41 & J162540.66-232811.3 & 246.4194 & -23.4698 & 9.06 & 0.02 & 9.02 & 0.02 & 8.70 & 0.05 & 5.31 & 0.05 & 10.45 & 0.02 & 9.53 & 0.02 & 9.25 & 0.02 & -24.00 & -13.00 & 1.40 &  III \\ 
   42 & J162609.65-232800.9 & 246.5402 & -23.4669 & 15.90 & 0.07 & 15.75 & 0.17 & 12.32 &  & 8.84 &  &  &  &  &  &  &  & 819.00 & 514.00 & 4.00 & \\ 
   46 & J162546.72-232737.0 & 246.4447 & -23.4603 & 9.57 & 0.02 & 9.45 & 0.02 & 9.60 & 0.06 & 8.80 &  & 11.02 & 0.02 & 10.09 & 0.03 & 9.77 & 0.02 & -82.00 & 18.00 & 0.90 & III \\ 
   47 & J162515.02-232713.8 & 246.3126 & -23.4538 & 10.74 & 0.02 & 10.46 & 0.02 & 10.31 & 0.15 & 7.92 &  & 12.15 & 0.02 & 11.30 & 0.02 & 10.93 & 0.02 & -164.00 & 26.00 & 1.60 & III \\ 
   48 & J162517.82-232713.1 & 246.3243 & -23.4537 & 9.07 & 0.02 & 8.94 & 0.02 & 8.81 & 0.05 & 8.19 &  & 10.46 & 0.02 & 9.54 & 0.02 & 9.25 & 0.02 & -114.00 & 87.00 & 1.50 & III \\ 
   50 & J162535.09-232648.8 & 246.3962 & -23.4469 & 3.52 & 0.38 & 3.19 & 0.24 & 3.55 & 0.02 & 3.52 & 0.02 & 3.57 & 0.29 & 3.35 & 0.25 & 3.17 & 0.52 & 554.00 & 846.00 & 1.00 &  \\ 
   51 & J162531.77-232646.1 & 246.3824 & -23.4462 & 9.54 & 0.02 & 9.29 & 0.02 & 9.41 & 0.07 & 7.14 & 0.14 &  &  &  &  &  &  & -35.00 & 105.00 & 1.10 & III \\ 
   52 & J162539.42-232642.2 & 246.4143 & -23.4451 & 9.13 & 0.02 & 8.64 & 0.02 & 6.79 & 0.02 & 5.04 & 0.04 & 10.96 & 0.02 & 10.03 & 0.03 & 9.57 & 0.02 & -18.00 & 21.00 & 5.30 & DD \\ 
   53 & J162532.53-232626.7 & 246.3856 & -23.4408 & 9.23 & 0.02 & 8.97 & 0.02 & 7.54 & 0.02 & 4.48 & 0.02 & 10.88 & 0.04 & 10.03 & 0.03 & 9.69 & 0.02 & -176.00 & 134.00 & 0.80 & DD \\ 
   55 & J162616.91-232605.5 & 246.5705 & -23.4349 & 16.14 & 0.09 & 15.02 & 0.10 & 11.66 & 0.34 & 8.39 &  &  &  &  &  &  &  & -513.00 & -110.00 & 1.90 & \\ 
   56 & J162620.73-232526.4 & 246.5864 & -23.4240 & 15.72 & 0.07 & 14.90 & 0.09 & 10.38 & 0.11 & 7.80 & 0.26 &  &  &  &  &  &  & -1701.00 & 851.00 & 5.80 & \\ 
   57 & J162542.88-232526.4 & 246.4287 & -23.4240 & 8.59 & 0.02 & 8.16 & 0.02 & 6.53 & 0.02 & 4.43 & 0.03 & 10.70 & 0.03 & 9.62 & 0.02 & 9.20 & 0.02 & -161.00 & 108.00 & 0.70 & DD \\ 
   58 & J162538.89-232523.7 & 246.4121 & -23.4233 & 9.95 & 0.02 & 9.71 & 0.02 & 9.64 & 0.06 & 8.74 &  & 11.39 & 0.02 & 10.56 & 0.03 & 10.17 & 0.02 & -30.00 & 46.00 & 2.50 & III \\ 
   61 & J162555.12-232424.7 & 246.4797 & -23.4069 & 9.42 & 0.02 & 9.27 & 0.02 & 9.05 & 0.04 & 8.16 & 0.29 & 10.99 & 0.03 & 9.95 & 0.02 & 9.64 & 0.02 & -136.00 & 25.00 & 0.90 & III \\ 
   62 & J162534.97-232417.5 & 246.3957 & -23.4049 & 6.29 & 0.06 & 6.12 & 0.04 & 5.90 & 0.02 & 5.66 & 0.05 & 6.11 & 0.02 & 5.92 & 0.04 & 5.78 & 0.02 & -1746.00 & 1090.00 & 3.20 & \\ 
   64 & J162538.17-232402.3 & 246.4090 & -23.4006 & 12.85 & 0.02 & 12.96 & 0.03 & 11.64 &  & 8.98 &  & 14.61 & 0.03 & 13.78 & 0.02 & 13.49 & 0.04 & 180.00 & -342.00 & 1.60 & \\ 
   66 & J162540.29-232352.6 & 246.4179 & -23.3980 & 12.28 & 0.02 & 12.03 & 0.02 & 11.37 &  & 8.90 &  & 13.51 & 0.03 & 12.85 & 0.03 & 12.50 & 0.03 & -144.00 & 80.00 & 2.60 & III \\ 
   67 & J162556.39-232346.7 & 246.4850 & -23.3963 & 15.66 & 0.07 & 14.92 & 0.10 & 11.17 & 0.24 & 8.41 &  &  &  &  &  &  &  & 412.00 & 589.00 & 2.90 & \\ 
   69 & J162545.84-232252.9 & 246.4410 & -23.3814 & 15.04 & 0.05 & 15.03 & 0.11 & 11.84 &  & 8.30 &  & 16.28 & 0.09 & 15.37 & 0.10 & 14.90 & 0.13 & -228.00 & 100.00 & 3.70 & \\ 
   71 & J162516.88-232203.4 & 246.3204 & -23.3676 & 8.80 & 0.02 & 8.70 & 0.02 & 8.59 & 0.04 & 7.13 & 0.19 & 10.19 & 0.02 & 9.32 & 0.02 & 9.02 & 0.02 & -35.00 & 162.00 & 2.40 & III \\ 
   72 & J162557.23-232158.8 & 246.4885 & -23.3664 & 12.62 & 0.02 & 12.47 & 0.03 & 11.06 & 0.33 & 8.05 & 0.36 & 13.75 & 0.03 & 13.11 & 0.03 & 12.79 & 0.03 & -217.00 & 238.00 & 1.50 & \\ 
   74 & J162502.36-232145.0 & 246.2599 & -23.3625 & 9.14 & 0.02 & 9.06 & 0.02 & 8.82 & 0.05 & 7.99 &  & 10.47 & 0.02 & 9.58 & 0.02 & 9.29 & 0.02 & -25.00 & 79.00 & 0.60 & III \\ 
   76 & J162516.84-232135.0 & 246.3202 & -23.3597 & 9.88 & 0.02 & 9.73 & 0.02 & 9.29 & 0.05 & 7.78 & 0.21 & 11.32 & 0.02 & 10.43 & 0.02 & 10.12 & 0.02 & -16.00 & 248.00 & 0.30 & \\ 
   78 & J162454.14-232104.7 & 246.2256 & -23.3513 & 15.49 & 0.07 & 14.21 & 0.06 & 11.35 & 0.34 & 8.21 & 0.40 &  &  &  &  &  &  & 402.00 & -726.00 & 2.30 & \\ 
   79 & J162454.90-232049.2 & 246.2288 & -23.3470 & 10.65 & 0.02 & 10.38 & 0.02 & 10.14 & 0.12 & 7.77 & 0.41 & 12.00 & 0.02 & 11.23 & 0.02 & 10.87 & 0.02 & -88.00 & 141.00 & 3.30 & III \\ 
   80 & J162525.36-232009.6 & 246.3557 & -23.3360 & 16.67 & 0.15 & 15.89 & 0.22 & 11.39 &  & 8.32 & 0.42 &  &  &  &  &  &  & -888.00 & 2084.00 & 1.00 & \\ 
   81 & J162528.54-232004.9 & 246.3689 & -23.3347 & 15.20 & 0.05 & 14.11 & 0.06 & 10.95 & 0.17 & 7.87 & 0.47 &  &  &  &  &  &  & -552.00 & 7.00 & 1.20 & \\ 
   83 & J162552.82-231936.5 & 246.4701 & -23.3268 & 7.67 & 0.03 & 7.56 & 0.02 & 7.43 & 0.02 & 7.51 & 0.23 & 9.26 & 0.02 & 8.21 & 0.04 & 7.87 & 0.03 & -67.00 & 173.00 & 1.50 & III \\ 
   84 & J162510.50-231914.7 & 246.2938 & -23.3208 & 6.95 & 0.06 & 6.55 & 0.02 & 4.74 & 0.02 & 2.25 & 0.02 & 9.13 & 0.02 & 8.07 & 0.03 & 7.51 & 0.03 & 54.00 & -12.00 & 1.30 & \\ 
   86 & J162554.44-231645.3 & 246.4769 & -23.2793 & 10.06 & 0.02 & 9.84 & 0.02 & 9.42 & 0.05 & 8.70 &  & 11.65 & 0.03 & 10.69 & 0.02 & 10.30 & 0.02 & -96.00 & 48.00 & 1.70 & III \\ 
   87 & J162603.70-231553.7 & 246.5154 & -23.2649 & 15.88 & 0.08 & 14.69 & 0.08 & 12.10 & 0.47 & 8.33 &  &  &  &  &  &  &  & 496.00 & 37.00 & 2.30 & \\ 
   88 & J162537.55-231537.3 & 246.4065 & -23.2604 & 16.34 & 0.13 & 15.34 & 0.14 & 11.57 & 0.40 & 7.39 &  &  &  &  &  &  &  & 390.00 & 401.00 & 2.70 & \\ 
   89 & J162516.66-231522.4 & 246.3194 & -23.2562 & 15.69 & 0.06 & 14.95 & 0.10 & 11.12 & 0.20 & 8.59 &  &  &  &  &  &  &  & 197.00 & -2933.00 & 4.70 & \\ 
   \hline
\end{tabular}
}
\end{sideways}
\end{center}
\end{table}

\section{Parameters of best fit modeling of X-ray spectra}
\begin{table}[t]
\caption{\label{tabfit} Best fit parameters from spectral analysis of bright X-ray sources.
Notes: for source nr. 41 we give  the best fit of the average spectrum, the flare spectrum and post-flare spectrum.
For source 62 we give the  best fit of the average spectrum, while detailed spectroscopy is given in Table \ref{fit_rhoOphC}. 
}
\begin{center}
\begin{sideways}
\resizebox{1.3\textwidth}{!}{
\begin{tabular}{lllrrrrrrrrrrrlrrrrr}
  \hline \hline
  Src. & Camera & Model & Chi2 & DOF & Nh & eNh & $kT_1$ & e$kT_1$ & $N_1$ & e$N_1$ & $kT_2$ & e$kT_2$ & N$_2$ & e$N_2$ & $kT_3$ & e$kT_3$ & $N_3$ & e$N_3$ & flux \\ 
       &        &       &      &     & $cm^{-2}$ & $cm^{-2}$ & keV & keV & cm$^{-5}$ &  cm$^{-5}$ & keV & keV & cm$^{-5}$ &  cm$^{-5}$ & keV & keV & cm$^{-5}$ &  cm$^{-5}$ & erg s$^{-1}$cm$^{-2}$ \\
  \hline
1 & EPIC & 1T & 56.32 &  44 & 0.36 & 0.03 & 2.5 & 0.21 & 8.00e-04 & 7.00e-05 & -- & -- & -- & -- & -- & -- & -- & -- & 8.7e-13 \\ 
  12N & EPIC & 1T & 83.89 &  44 & 0.22 & 0.10 & 0.95 & 0.05 & 4.60e-05 & 1.00e-05 & -- & -- & -- & -- & -- & -- & -- & -- & 5.7e-14 \\ 
  17 & EPIC & 2T & 105.20 &  77 & 0.43 & 0.04 & 0.85 & 0.05 & 1.00e-04 & 2.50e-05 & 1.8 & 0.11 & 2.40e-04 & 1.50e-05 & -- & -- & -- & -- & 3.9e-13 \\ 
  20 & MOS & 1T & 91.20 &  59 & 0.33 & 0.04 &   1 & 0.04 & 7.01e-04 & 6.40e-05 & -- & -- & -- & -- & -- & -- & -- & -- & 5e-13 \\ 
  26 & EPIC & 1T & 35.50 &  29 & 0.29 & 0.05 &   1 & 0.05 & 8.00e-05 & 1.30e-05 & -- & -- & -- & -- & -- & -- & -- & -- & 6.7e-14 \\ 
  30 & EPIC & 1T & 12.70 &  19 & 0.31 & 0.09 & 0.88 & 0.08 & 6.00e-05 & 2.00e-05 & -- & -- & -- & -- & -- & -- & -- & -- & 5.8e-14 \\ 
  31 & EPIC & 3T & 178.20 & 133 & 0.50 & 0.06 & 0.23 & 0.02 & 1.60e-03 & 1.00e-03 & 1.1 & 0.04 & 1.00e-03 & 2.00e-04 & 2.1 & 0.56 & 2.50e-04 & 1.40e-04 & 1.8e-12 \\ 
  38 & EPIC & 2T & 66.70 &  61 & 0.55 & 0.08 & 0.24 & 0.03 & 8.20e-04 & 5.20e-04 &   1 & 0.07 & 2.08e-04 & 3.00e-05 & -- & -- & -- & -- & 5.2e-13 \\ 
  41 & EPIC & 2T & 115.00 &  90 & 0.20 & 0.30 & 0.99 & 0.03 & 8.10e-04 & 8.00e-06 & 2.4 & 0.21 & 1.10e-04 & 1.10e-05 & -- & -- & -- & -- & 2.3e-13 \\ 
  41 Flare & EPIC & 2T & 49.70 &  29 & 0.20 & -- & 0.93 & -- & 7.20e-05 & -- & 2.5 & 0.21 & 3.10e-04 & 2.00e-05 & -- & -- & -- & -- & 4.6e-13 \\ 
  41 Post-Flare & EPIC & 2T & 77.04 &  64 & 0.20 & 0.30 & 0.93 & 0.04 & 7.20e-05 & 1.00e-05 & 2.2 & 0.23 & 9.40e-05 & 1.10e-05 & -- & -- & -- & -- & 2e-13 \\ 
  46 & EPIC & 1T & 40.90 &  28 & 0.29 & 0.06 & 0.93 & 0.05 & 5.00e-05 & 6.00e-06 & -- & -- & -- & -- & -- & -- & -- & -- & 6.1e-13 \\ 
  48 & EPIC & 2T & 62.20 &  57 & 0.27 & 0.06 & 0.99 & 0.05 & 1.18e-04 & 3.00e-05 & 2.5 & 0.4 & -- & -- & -- & -- & -- & -- & 3.3e-13 \\ 
  50 & EPIC & 3T & 252.60 & 207 & 0.48 & 0.04 & 0.24 & 0.02 & 2.28e-03 & 9.00e-04 &   1 & 0.02 & 1.64e-03 & 2.00e-04 & 3.4 & 0.72 & 3.90e-04 & 1.00e-04 & 3e-12 \\ 
  52 & MOS & 1T & 14.70 &  10 & 0.63 & 0.24 & 1.1 & 0.24 & 8.00e-05 & 3.00e-05 & -- & -- & -- & -- & -- & -- & -- & -- & 9e-14 \\ 
  58 & PN & 1T & 1.55 &   4 & 0.65 & 0.13 & 0.6 & 0.12 & 5.00e-05 & 2.40e-05 & -- & -- & -- & -- & -- & -- & -- & -- & 5.9e-14 \\ 
  61 & PN & 1T & 57.10 &  39 & 0.42 & 0.03 & 2.4 & 0.17 & 1.30e-03 & 1.20e-04 & -- & -- & -- & -- & -- & -- & -- & -- & 1.4e-12 \\ 
  62 & EPIC & 3T & 265.60 & 225 & 0.52 & 0.04 & 0.3 & 0.03 & 1.80e-03 & 8.00e-04 & 1.1 & 0.02 & 2.05e-03 & 2.10e-04 & 5.4 & 1.4 & 4.20e-04 & 8.00e-05 & 3.1e-12 \\ 
  71 & PN & 3T & 41.00 &  39 & 0.40 & 0.08 & 0.24 & 0.05 & 5.30e-04 & 5.50e-04 &   1 & 0.09 & 3.60e-04 & 2.00e-04 & 1.9 & 0.4 & 2.10e-04 & 1.30e-04 & 6.9e-13 \\ 
  74 & EPIC & 1T & 28.10 &  22 & 0.56 & 0.09 & 0.93 & 0.08 & 1.20e-04 & 2.20e-05 & -- & -- & -- & -- & -- & -- & -- & -- & 1.5e-13 \\ 
  76 & EPIC & 1T & 22.00 &  16 & 0.64 & 0.07 & 0.75 & 0.06 & 1.50e-04 & 3.00e-05 & -- & -- & -- & -- & -- & -- & -- & -- & 1.9e-13 \\ 
  79 & EPIC & 1T & 1.40 &   1 & 0.06 & 0.05 & 1.3 & 0.13 & 1.33e-05 & 3.00e-06 & -- & -- & -- & -- & -- & -- & -- & -- & 1.5e-14 \\ 
  83 & EPIC & 2T & 78.40 &  69 & 0.49 & 0.04 &   1 & 0.04 & 7.10e-04 & 2.30e-04 & 2.9 & 1.4 & 1.60e-04 & 1.00e-04 & -- & -- & -- & -- & 7.7e-13 \\ 
  84 & EPIC & 2T & 108.70 &  84 & 0.35 & 0.02 & 0.83 & 0.04 & 3.00e-04 & 7.00e-05 & 1.6 & 0.1 & 3.30e-04 & 4.00e-05 & -- & -- & -- & -- & 6e-13 \\ 
\hline \hline
  Src. & Camera & Model & Chi2 & DOF & $N_H$       &       e$N_H$ & $\alpha$ & error & Norm     & e.Norm      &\multicolumn{8}{c}{ } & flux \\ 
       &	&       &      &     & \multicolumn{2}{c}{cm$^{-2}$} &          &       & \multicolumn{2}{c}{cm$^{-5}$} & \multicolumn{8}{c}{ } & erg s$^{-1}$cm$^{-2}$ \\
   \hline
  12S & EPIC & Pow & 83.89 &  44 & 0.45 & 0.30 & 1.8 & 0.23 & 2.20e-05 & 7.00e-06 & -- & -- & -- & -- & -- & -- & -- & -- & 1.3e-13 \\ 
  44 & EPIC & Pow & 27.20 &  22 & 0.52 & 0.10 & 1.5 & 0.15 & 6.00e-05 & 1.00e-05 & -- & -- & -- & -- & -- & -- & -- & -- & 4.2e-13 \\ 
  54 & MOS & Pow & 36.00 &  25 & 0.71 & 0.13 & 1.4 & 0.15 & 5.50e-05 & 1.00e-05 & -- & -- & -- & -- & -- & -- & -- & -- & 4.5e-13 \\ 
  78 & EPIC & Pow & 1.73 &   2 & 0.33 & 0.16 & 1.6 & 0.33 & 8.00e-06 & 3.00e-06 & -- & -- & -- & -- & -- & -- & -- & -- & 5.7e-14 \\ 
  81 & EPIC & Pow & 27.34 &  31 & 0.71 & 0.10 & 1.7 & 0.12 & 4.00e-05 & 6.00e-06 & -- & -- & -- & -- & -- & -- & -- & -- & 2.5e-13 \\ 
   \hline
\end{tabular}}
\end{sideways}
\end{center}
\end{table}
\section{Results from best fit to SEDs}
\begin{table}[t]
\centering
\caption{\label{vosafit} Parameters of bestfit to SEDs.} 
\begin{center}
\begin{sideways}
\resizebox{1.3\textwidth}{!}{
\begin{tabular}{cccccccccccccccc}
  \hline
  \hline
Source & Model & $T_\mathrm{eff}$ & $e(T_\mathrm{eff})$ & $\log g$ & $e(\log g)$ & Metallicity & $e(\mathrm{Met.})$ & F$_\mathrm{tot}$ & $e(\mathrm{F_{tot})}$ & $L_\mathrm{bol}$ & $e(L_\mathrm{bol})$ & Mass & Age\\ 
       &       & K    &   K     & cm s$^{-2}$ &  cm s$^{-2}$ &       &      & erg s$^{-1}$cm$^2$ & erg s$^{-1}$cm$^2$ & $L_\mathrm{bol,\odot}$ &  $L_\mathrm{bol,\odot}$ & M$_\odot$ & Myr \\
  \hline
  1 & bt-settl-cifist & 3300 & 61 & 4.0 & 0.3 & 0.0 & 0.0 & 2.45e-10 & 8.1e-12 & 1.10e-01 & 3.6e-03 & 0.26 & 3.0 \\ 
   12 & cond00 & 4000 & 38 & 4.5 & 0.3 & 0.0 & 0.0 & 5.57e-10 & 1.4e-11 & 2.50e-01 & 6.4e-03 & 0.88 & 20.0 \\ 
   16 & cond00 & 3000 & 93 & 3.5 & 0.4 & 0.0 & 0.0 & 5.49e-11 & 2.3e-12 & 2.46e-02 & 1.0e-03 & 0.10 & 5.2 \\ 
   20 & bt-nextgen-agss2009 & 5100 & 119 & 1.0 & 0.7 & -1.0 & 0.5 & 1.87e-09 & 1.4e-10 & 8.38e-01 & 6.4e-02 & --- & --- \\ 
   26 & bt-nextgen-gns93 & 3600 & 71 & 4.0 & 0.4 & 0.0 & 0.2 & 3.23e-10 & 1.7e-11 & 1.45e-01 & 7.5e-03 & --- & --- \\ 
   30 & dusty00 & 3700 & 88 & 4.5 & 0.2 & --- & --- & 3.45e-10 & 1.7e-11 & 1.55e-01 & 7.7e-03 & 0.68 & 14.4 \\ 
   31 & bt-settl & 4800 & 112 & -0.5 & 0.2 & -1.5 & 0.7 & 2.60e-09 & 1.9e-10 & 1.17e+00 & 8.6e-02 & 1.20 & --- \\ 
   38 & cond00 & 3500 & 58 & 4.5 & 0.2 & 0.0 & 0.0 & 5.06e-10 & 1.9e-11 & 2.27e-01 & 8.6e-03 & 0.46 & 3.5 \\ 
   41 & bt-nextgen-agss2009 & 4300 & 89 & -0.5 & 0.1 & -1.0 & 0.6 & 7.88e-10 & 4.6e-11 & 3.54e-01 & 2.0e-02 & --- & --- \\ 
   46 & Kurucz & 3750 & 76 & 5.0 & 0.2 & 0.5 & 0.2 & 4.18e-10 & 1.9e-11 & 1.88e-01 & 8.5e-03 & 0.50 & 7.1 \\ 
   47 & dusty00 & 3300 & 68 & 4.0 & 0.4 & --- & --- & 1.14e-10 & 5.3e-12 & 5.14e-02 & 2.4e-03 & 0.25 & 8.7 \\ 
   48 & bt-settl-cifist & 3700 & 86 & 4.5 & 0.3 & 0.0 & 0.0 & 5.92e-10 & 3.3e-11 & 2.66e-01 & 1.5e-02 & 0.50 & 3.0 \\ 
   52 & cond00 & 3100 & 113 & 4.5 & 0.6 & 0.0 & 0.0 & 4.71e-10 & 2.4e-11 & 2.11e-01 & 1.1e-02 & --- & --- \\ 
   57 & bt-settl-cifist & 4200 & 105 & 5.5 & 0.3 & 0.0 & 0.0 & 9.72e-10 & 6.0e-11 & 4.36e-01 & 2.7e-02 & 0.90 & 7.5 \\ 
   58 & dusty00 & 3300 & 46 & 4.0 & 0.2 & --- & --- & 2.38e-10 & 8.7e-12 & 1.07e-01 & 3.9e-03 & 0.28 & 3.7 \\ 
   61 & bt-nextgen-agss2009 & 3600 & 67 & 4.0 & 0.4 & 0.0 & 0.3 & 3.98e-10 & 1.6e-11 & 1.79e-01 & 7.1e-03 & --- & --- \\ 
   66 & Kurucz & 3500 & 0 & 5.0 & 0.0 & 0.0 & 0.2 & 3.59e-11 & 9.1e-13 & 1.61e-02 & 4.1e-04 & 0.30 & 90.4 \\ 
   71 & bt-settl-cifist & 3900 & 102 & 5.5 & 0.3 & 0.0 & 0.0 & 8.38e-10 & 5.0e-11 & 3.76e-01 & 2.2e-02 & 0.62 & 3.4 \\ 
   74 & bt-nextgen-gns93 & 4000 & 112 & 5.0 & 0.3 & 0.3 & 0.1 & 6.27e-10 & 4.3e-11 & 2.82e-01 & 1.9e-02 & --- & --- \\ 
   79 & dusty00 & 3300 & 91 & 5.0 & 0.6 & --- & --- & 1.25e-10 & 6.5e-12 & 5.61e-02 & 2.9e-03 & 0.25 & 7.9 \\ 
   86 & cond00 & 3100 & 50 & 4.5 & 0.2 & 0.0 & 0.0 & 1.66e-10 & 3.0e-12 & 7.44e-02 & 1.4e-02 & 0.16 & 2.1 \\ 
\hline
\end{tabular}
}
\end{sideways}
\end{center}
\end{table}

\section{Light curves}

\begin{figure*}
\begin{center}
\resizebox{\textwidth}{!}{\includegraphics{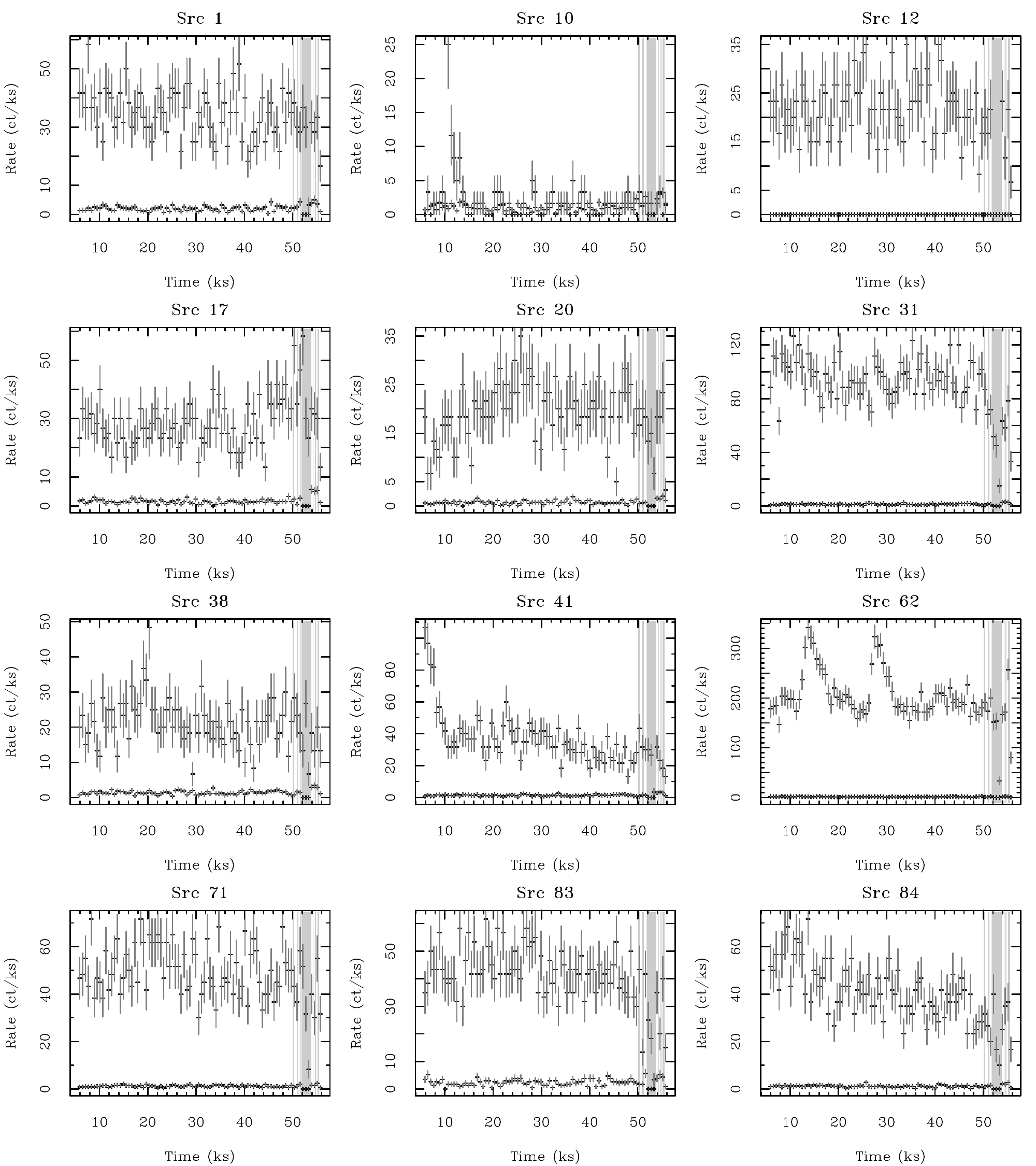}}
\end{center}
\caption{\label{lcs} PN light curves of brightest X-ray sources and { showing variability at $>1 \sigma$ level.}
Light curves of scaled background are also shown. 
Sources 10, 38, 41 and 62 show clear flare-like variability. 
Gray areas represent the high background intervals filtered out for optimizing the
source detection effectiveness toward faint sources.
Source 62 (\rhoOph~C) is discussed in details in Sect. \ref{rhoophC}. The light curve of source 50 
(\rhoOph~A+B) is not shown here, its analysis is detailed in \citet{Pillitteri2014c}.}
\end{figure*}

\end{document}